\documentclass[aps,pra,twocolumn,showpacs,preprintnumbers,amsmath,amssymb]{revtex4-1}
\usepackage{ifpdf}
 \newif\ifpdf
\ifx\pdfoutput\undefined
   \pdffalse
\else
   \pdfoutput=1
   \pdftrue
\fi
\ifpdf
   \usepackage{graphicx}
   \usepackage{epstopdf}
   \usepackage{color}
\else
   \usepackage{graphicx}
\fi

\usepackage{tabulary}
\usepackage{srctex}%\usepackage{hyperref}
\usepackage{hyperref}

\DeclareMathOperator{\mum}{{\mu m}}
\DeclareMathOperator{\OR}{\Omega_{\mathrm{\scriptscriptstyle R}}}
\DeclareMathOperator{\epsD}{\varepsilon_{\mathrm{\scriptscriptstyle D}}}
\DeclareMathOperator{\kappaD}{\kappa_{\mathrm{\scriptscriptstyle D}}}
\DeclareMathOperator{\kB}{\mathit k_{\mathrm{\scriptscriptstyle B}}}
\DeclareMathOperator{\Ef}{\mathit E_{\mathrm{\scriptscriptstyle F}}}
\DeclareMathOperator{\Tr}{\mathrm{Tr}}
\begin{document}
\title{Self-consistent Description of Graphene Quantum Amplifier}

\author{Yu.E. Lozovik}
\affiliation{Institute for Spectroscopy RAS, 5 Fizicheskaya, Troitsk 142190, Russia}
\affiliation{Moscow Institute of Physics and Technology, 9 Institutskiy pereulok, Dolgoprudny 141700, Moscow Region, Russia}
\affiliation{All-Russia Research Institute of Automatics, 22 Suschevskaya, Moscow 127055, Russia}
\affiliation{Moscow National Research Nuclear University MEPhI, 31 Kashirskoe highway, Moscow 115409, Russia}

\author{I.A. Nechepurenko}
\affiliation{All-Russia Research Institute of Automatics, 22 Suschevskaya, Moscow 127055, Russia}
\affiliation{Moscow Institute of Physics and Technology, 9 Institutskiy pereulok, Dolgoprudny 141700, Moscow Region, Russia}

\author{ A.V. Dorofeenko}
\affiliation{All-Russia Research Institute of Automatics, 22 Suschevskaya, Moscow 127055, Russia}
\affiliation{Moscow Institute of Physics and Technology, 9 Institutskiy pereulok, Dolgoprudny 141700, Moscow Region, Russia}
\affiliation{Institute for Theoretical and Applied Electromagnetics, 13 Izhorskaya, Moscow 125412, Russia}

\author{E.S. Andrianov}
\affiliation{All-Russia Research Institute of Automatics, 22 Suschevskaya, Moscow 127055, Russia}
\affiliation{Moscow Institute of Physics and Technology, 9 Institutskiy pereulok, Dolgoprudny 141700, Moscow Region, Russia}

\author{N.M. Chtchelkatchev}
\affiliation{L.D. Landau Institute for Theoretical Physics, Russian Academy of Sciences, 117940 Moscow, Russia}
\affiliation{Moscow Institute of Physics and Technology, 141700 Moscow, Russia}
\affiliation{Institute for High Pressure Physics, Russian Academy of Sciences, 142190 Troitsk, Russia}

\author{A.A Pukhov}
\affiliation{All-Russia Research Institute of Automatics, 22 Suschevskaya, Moscow 127055, Russia}
\affiliation{Moscow Institute of Physics and Technology, 9 Institutskiy pereulok, Dolgoprudny 141700, Moscow Region, Russia}
\affiliation{Institute for Theoretical and Applied Electromagnetics, 13 Izhorskaya, Moscow 125412, Russia}

\date{\today}
\begin{abstract}
High level of dissipation in \textit{normal} metals makes challenging development of active and passive plasmonic devices. One possible solution to this problem is to use \textit{alternative} materials. Graphene is a good candidate for plasmonics in near infrared (IR) region. In this paper we develop quantum theory of a graphene plasmon generator.  We account for the first time  quantum correlations and dissipation effects that allows describing such regimes of quantum plasmonic amplifier as surface plasmon emitting diode and surface plasmon amplifier by stimulated emission of radiation.  Switching between these generation types is possible \textit{in situ} with variance of graphene Fermi-level or gain transition frequency. We provide explicit expressions for dissipation and interaction constants through material parameters and find the generation spectrum and correlation function of second order which predicts laser statistics.
\end{abstract}

\maketitle

\section{Introduction}

The recent developments of plasmonics~\cite{maier2007plasmonics,shvets2011plasmonics,enoch2012plasmonics,zouhdi2008metamaterials,bozhevolnyiplasmonic,shalaev2006nanophotonics,zayats2013active,sarychev2007electrodynamics,west2010LPR,Lozovik1987} made possible the creation of plasmonic devices analogous to those in classical optics~\cite{Liu2007PhysRevLett}. Theoretical and experimental studies of plasmonic lenses, mirrors, and cavities have been performed~\cite{barnes2003Nature,radko2008OptExp,gong2007APL,archambault2009PRB,feng2007APL,zia2007NatNano}. Some important benefits of plasmonic devices over optical ones are their subwavelength focusing ability and high field intensity leading to strong field-matter interaction. Surface plasmons are important in the field of surface-enhanced spectroscopy. A high localization of plasmons increases sensitivity of the absorption spectroscopy to molecules located at the surface~\cite{pockrand1978surface,mills1982surface,mulvaney1996Lengmuir,eberlein2008PRB,tanaka2013development}. The latter effect also contributes to surface and tip enhancement of Raman scattering (SERS and TERS)~\cite{jeanmaire1977surface,otto1978SurfSci,eesley1981PRB,tsang1979PRL} which has made possible the detection of single molecules~\cite{fang2013RSC} and development of revolutionary diagnostic methods and promising bioanalysis applications in medicine and biology~\cite{Cao2002science,Ang2015hemoglobin}.

Plasmonics applications are limited by Ohmic losses in metal. The use of an active medium has been proposed recently for loss compensation~\cite{hawrylak1986APL,kempa1988} and amplification~\cite{maier2006optcomm} of surface plasmon-polaritons (SPPs) propagating along active nanostructures. The amplification can lead to SPP generation~\cite{sirtori1998OptLett,tredicucci2000APL,babuty2010PRL,oulton2009nature,noginov2008PRL}. Further, it has been understood that a surface plasmon localized at a single nanoparticle can also be coherently generated by radiationless excitation  by neighboring excited system~\cite{bergman2003PRL,protsenko2005PRA,andrianov2012PRB,andrianov2011OptExp,andrianov2011OptLett}. The experimental realization of such a system, spaser, was reported by several groups~\cite{noginov2009Nature,lu2012Science,hill2009OptExp,suh2012NanoLett,van2013PRL}. In general, the difference between the SPP generator (``SPP laser'') and spaser is vague, so that they are often identified~\cite{berini2012NatPhot}. On the other hand, these two devices can be considered as opposed limiting cases, respectively, of large SPP cavity with quasi continuous spectra and small (nanoscopic) system with discrete spectra. Beside a rich perspective in applications, plasmonic generators are interesting by themselves as pioneering devices of quantum plasmonics~\cite{tame2013NatPhys,jacob2011Science}.

One of the most promising material for plasmonics is graphene~\cite{ju2011NatNano,koppens2011NanoLett,grigorenko2012NatPhot,garcia2014ACSPhoton,low2014ACSNano,maier2012NatPhys}. It is an extremely thin 2D material~\cite{novoselov2005Nature,zhang2005Nature}, which has high carrier mobility~\cite{bolotin2008SolStCom}. This material supports plasmons with small damping. Unique property of graphene is \textit{in situ} control of electron Fermi level achievable not only  by doping as in typical semiconductor, but with the use of a gate electrode~\cite{balandin2008NanoLett}. That makes graphene the state-of-the-art material applicable from optical to THz region. The latter frequency region is of high interest for applications as it contains vibrational transitions of molecules. The use of graphene opens up opportunities to create highly sensitive compact THz devices. In near-IR graphene plasmons, the localization factor reaches higher values while losses are lower than in metal plasmons (that are typically in optic frequency range)~\cite{hwang2007PRB}. In spite of the relatively low absorption in graphene, the loss limit of the plasmon mean free path is still a major obstacle for graphene plasmonics applications. Recently active graphene plasmonics has become  rapidly developing~\cite{otsuji2014JPhysD}. In particular, graphene plasmon generators based on the use of a active medium have been proposed~\cite{berman2013PRB,rupasinghe2014ACSNano,apalkov2014LSA}.

However, the question about coherent properties of plasmons on graphene is opened. Semiclassical approximations which are usually used to investigate plasmon generation are hardly applicable for investigation of plasmon correlation function and spectrum because they do not take into account spontaneous transitions in active medium and quantum fluctuations. However these processes may break down signal coherence.

To solve this problem we use quantum master equation for field and active medium which takes into account all dissipation processes and quantum fluctuations. We account for the first time  quantum correlations and dissipation effects that allows to describe such regimes of quantum plasmonic amplifier as surface plasmon emitting diode (SPED) and surface plasmon amplification by stimulated emission of radiation (SPASER). We show that switching between these generation types is possible with variance of graphene Fermi-level or gain transition frequency. We find the generation spectrum and correlation function of second order which predicts laser statistics.

\section{Master equation of graphene nanolaser dynamics}
\begin{figure}[t]
  \centering
  \includegraphics[width=\columnwidth]{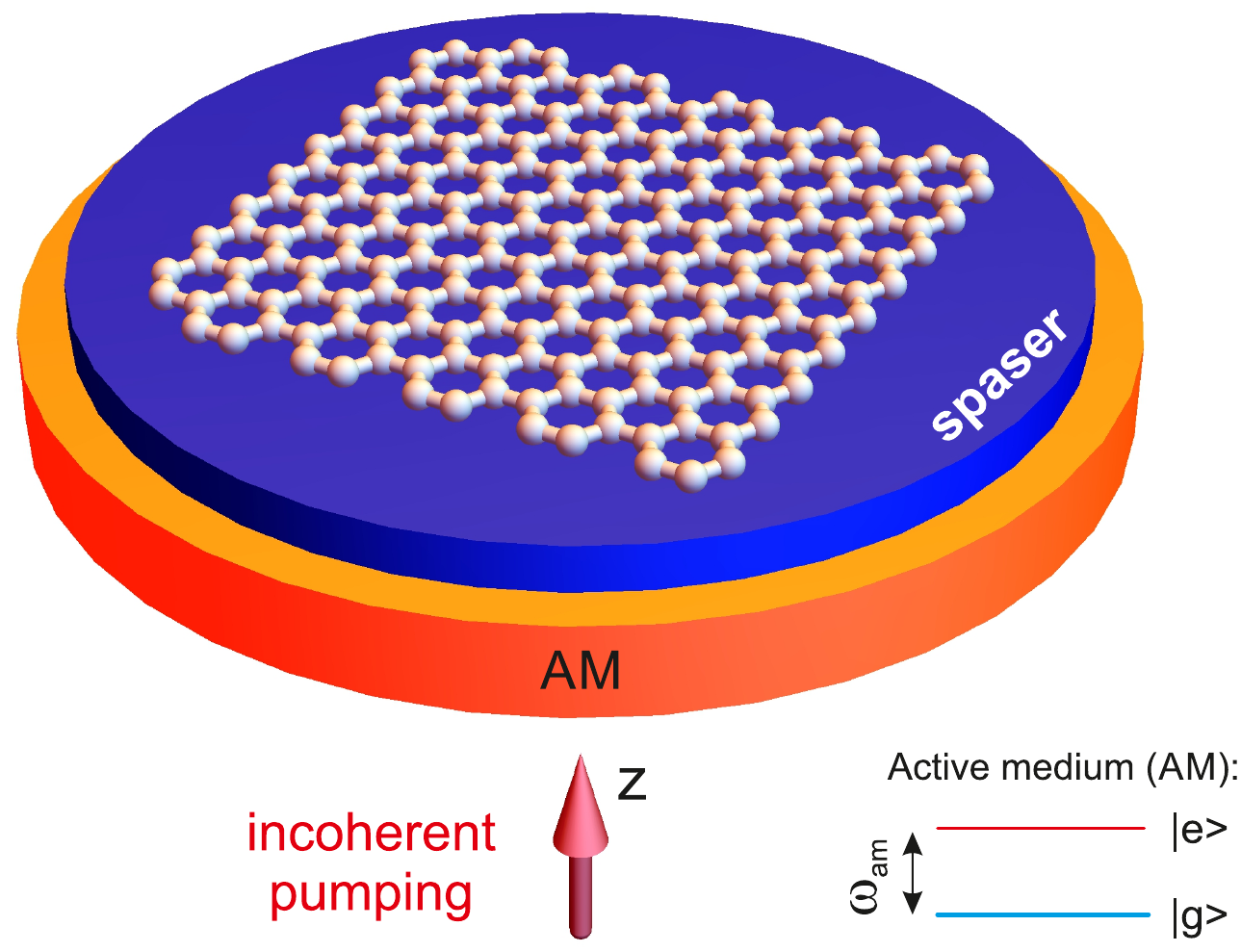}\\
  \caption{(Color online) A sketch of graphene quantum amplifier.}\label{fignanolaser}
\end{figure}

In the present paper graphene quantum amplifier, see Fig.~\ref{fignanolaser}, is consistently described. It is considered as an open quantum system. Below the graphene layer there is active medium (AM) layer with active molecules pumped incoherently. In Fig.~\ref{fignanolaser} we schematically show  the working two levels of the active molecule. Using quantum model we explore below how principal characteristics of the quantum amplifier develop with tuning of controlling parameters.

For simplicity we suppose that each particle or QD of active medium is multi-level system with ``working'' transition frequency $\omega_{\rm am}\sim\omega_{\rm pump}$. Also we assume that active medium with frequency $\omega_{\rm am} $ interacts only with one mode of surface plasmon with the wavelength $\lambda_{\rm pl} =2\pi c/\omega_{\rm am} $. After second quantization procedure plasmon energy spectrum are harmonic oscillator Fock states ${\left| n \right\rangle} $, where $n$ is number of excited plasmons.

If we focus on the working frequency then each active medium active atom can be effectively described by the two-level system with excited ${\left| e \right\rangle} $ and ground ${\left| g \right\rangle} $ states, see Fig.~\ref{fignanolaser}.

Graphene plasmons are investigated in the range $\lambda \sim 1-10\mum$ so the mean value of thermal boson number in reservoirs $n_{\rm th} =\left(\exp \left(\hbar \omega /k_{\rm B}T\right)-1\right)^{-1} \ll 1$ at room temperature. So the system dynamics in the Markovian approximation can be described by the Lindblad equation~\cite{carmichael2008book2}:
\begin{widetext}
\begin{multline} \label{GrindEQ__1_}
\frac d{dt}{\hat \rho }=-\frac{i}{\hbar } \left[\hat{H},\hat \rho \right]+\frac{\gamma_{\rm pl} }{2} \left(2\hat a\hat \rho \hat a^{+} -\hat a^{+} \hat a\hat \rho -\hat \rho \hat a^{+}\hat  a\right)+\sum_{k}\frac{\gamma_{\rm am}^{\rm decay} }{2} \left(2\hat \sigma_{k} \hat \rho \hat \sigma_{k}^{+} -\hat \sigma_{k}^{+} \hat \sigma_{k}\hat  \rho -\hat \rho \hat \sigma_{k}^{+}\hat  \sigma_{k} \right) +
\\
+\sum_{k}\frac{\gamma_{\rm am}^{\rm dephasing} }{2} \left(\hat \sigma_{z,k}\hat  \rho \hat \sigma_{z,k} -\hat \rho \right) +\sum_{k}\frac{\gamma_{\rm am}^{\rm pump} }{2} \left(2\hat \sigma_{k}^{+} \hat \rho \hat \sigma_{k} -\hat \sigma_{k} \hat \sigma_{k}^{+}\hat  \rho -\hat \rho\hat  \sigma_{k}\hat  \sigma_{k}^{+} \right).
\end{multline}
\end{widetext}
Here $\hat \rho $ is the density matrix of the whole ``plasmon + active medium'' system which Hilbert space is the direct product of the plasmon and two-level atoms Hilbert spaces. The first term in the right-hand side of \eqref{GrindEQ__1_} describes hermitian Hamiltonian dynamics, where  $\hat H=\hbar \omega_{\rm pl} \left(\hat{a}^{+} \hat{a}+\frac{1}{2} \right)+\sum_{k}\hbar \omega _{\rm am} \hat{\sigma }^{+} \hat{\sigma } +\sum_{k}\hbar \OR \left(\hat{a}^{+} \hat{\sigma }_{k} +\hat{a}\hat{\sigma }_{k}^{+} \right)$ consists of three contributions:  Hamiltonian of plasmon, two-level system and the interaction between them. Here $\hat{a}$ and $\hat{a}^{+} $ are plasmon creation and annihilation operators, respectively, $\hat{\sigma }_{k} ={\left| g_{k}  \right\rangle} {\left\langle e_{k}  \right|} $ and $\hat{\sigma }_{k}^{+} ={\left| e_{k}  \right\rangle} {\left\langle g_{k}  \right|} $ are the transition operators of $k$-th two-level atom, $\Omega_{R} $ is the Rabi constant of the interaction between plasmon and active medium, $\omega_{\rm pl} =2\pi c/\lambda_{pl} $ is the surface plasmon frequency. As we mentioned above during calculation we suppose $\omega_{\rm pl} =\omega_{\rm am} $. Second term is the Lindblad part corresponding to the plasmon damping with the rate $\gamma_{\rm pl} $. Third and fourth terms correspond to energy decay with the rate $\gamma_{\rm am}^{\rm decay} $ and dephasing with the rate $\gamma_{\rm am}^{\rm dephasing} $ (also known as longitudinal and transverse relaxations). Finally, last term is responsible for incoherent pumping of two-level atom with the rate $\gamma_{\rm am}^{\rm pump} $.

Note that we neglect interaction between two-level particles. This may be justified when interaction between each atom and surface plasmon is much larger than interaction between atoms. If so, interaction between atoms results in energy dissipation and redifinition of pumping rates.
\begin{figure}[t]
\includegraphics[width=\columnwidth]{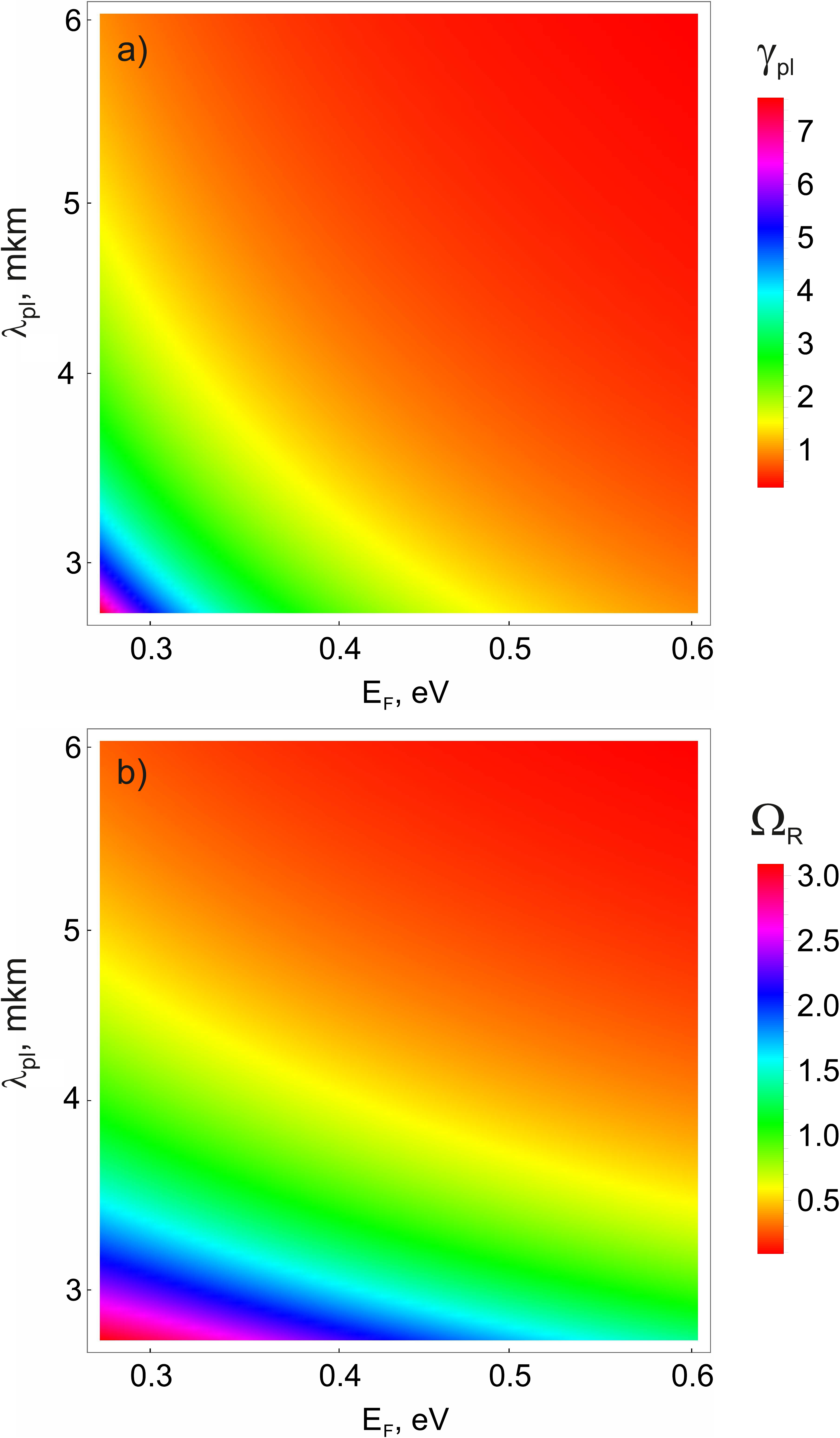}
\caption{(Color online) (a) The dependencies of the damping rate of surface plasmon $\gamma_{\rm pl} $ and (b) Rabi constant of interaction between plasmon and active medium $\OR $ on Fermi level $E_F$ and plasmon wavelength $\lambda_{\rm pl} $.\label{gammaPlOmegaR}}
\end{figure}

Suppose that there is no frequency mismatch between atomic dipole moments. This is true when all atoms occupy subwavelength volume or distance between them is of the order of plasmon wavelength. So all atoms may be considered as identical and it is possible to introduce collective atomic operator $\hat{J}=\sum _{k}\hat{\sigma }_{k}  $, $\hat{J}^{+} =\sum_{k}\hat{\sigma }_{k}^{+}  $, and $\hat{J}_{z} =\sum _{k}\hat{\sigma }_{z,k}  $. With the aid of these operators and the assumption that all atoms are identical it is possible to rewrite Eq.~\eqref{GrindEQ__1_} in the form
\begin{multline} \label{GrindEQ__2_}
\frac d{dt}{\hat \rho }=-\frac{i}{\hbar } \left[\hat{H},\hat \rho \right]+\frac{\gamma_{\rm pl} }{2}\mathcal L[\hat a,\hat a^+]+
\frac{\gamma_{\rm am}^{\rm decay} }{2}\mathcal L[\hat J,\hat J^+]+
\\
\frac{\gamma_{\rm am}^{\rm dephasing} }{2} \mathcal L[\hat J_z,\hat J_z^+]+
\frac{\gamma_{\rm am}^{\rm pump} }{2} \mathcal L[\hat J^+,\hat J],
\end{multline}
where the Lindbladian, $\mathcal L[\hat A,\hat A^+]=2A\hat \rho \hat A^{+} -\hat A^{+} \hat A\hat \rho -\hat \rho \hat A^{+} \hat A$.

The form of Eq. \eqref{GrindEQ__2_} is standard, but important are specific values of the damping and pumping rates. The damping rates of surface plasmon and Rabi constant are controlling parameters. The value of damping constant of semiconductor active medium may be evaluated as follows: $\gamma_{\rm am}^{\rm decay} \sim 10^{11} s^{-1} $, $\gamma_{\rm am}^{\rm dephasing} \sim 10^{12} s^{-1} $~\cite{khurgin2012OE,khurgin2012APL}. The value of the pumping rate is experimentally controllable parameter. Its maximal value strongly depends on the type of active medium. In Refs.~\cite{khurgin2012OE,khurgin2012APL} it has been shown that the pumping rate $\gamma_{\rm am}^{\rm pump} \sim 10^{13} s^{-1} $ corresponds to the current density $\sim 10kA/cm^{2} $ which is nearly the maximal achievable value of current density today in semiconductors. For colloidal quantum dots and dye molecles which are pumped by external electromagnetic field corresponding field intensity is $E \simeq 10^{5} V/m$ ~\cite{andrianov2013loss}. We take in our model the value, $\gamma_{\rm am}^{\rm pump} \sim 10^{13} s^{-1} $, as maximum possible pumping rate .

\section{Numerical estimation of the main parameters, Rabi constant ($\OR$) and plasmon decay rate ($\gamma_{\rm pl} $)}

\subsection{ Rabi constant $\OR$}
\begin{figure*}[t]
\includegraphics[width=\textwidth]{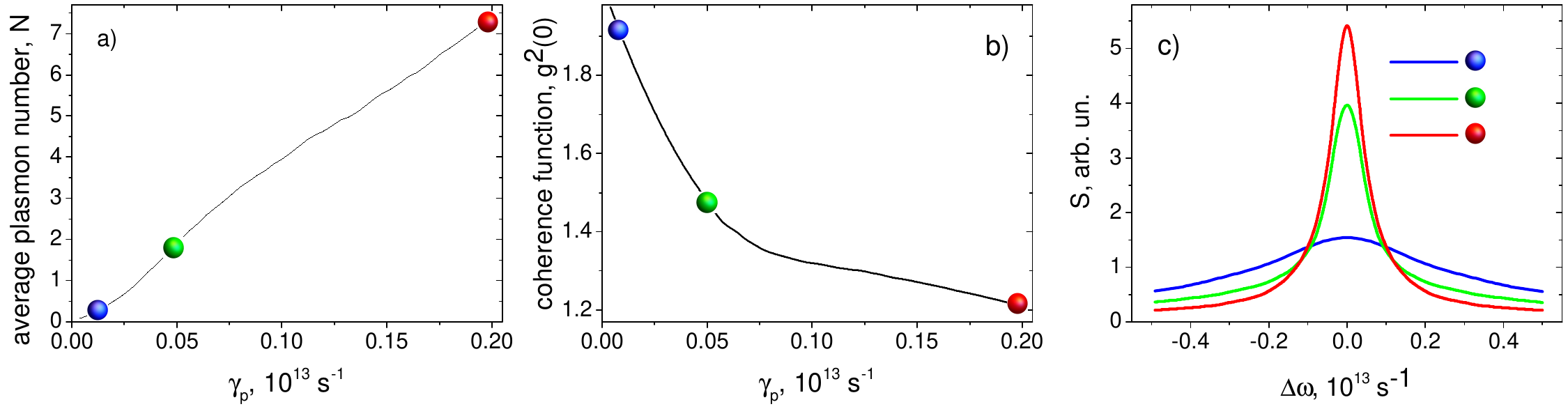}
\caption{(Color online) From left to right (a-c): the dependence of average number of excited plasmons $N$ on the pumping rate $\gamma_p$, the dependence of second order correlation function $g^{(2)}$ on the pumping rate $\gamma_p$ and spectrum of the plasmon field $S(\omega)$ at different values of pumping rates. Color balls, \textcolor{blue}{$\bullet$}, \textcolor{green}{$\bullet$}, and, \textcolor{red}{$\bullet$}, correspond to pumping rates at which spectra have been calculated. Parameters of graphene are the following: $\Ef \simeq 0.5eV$, $\lambda _{\rm pl} =5\mum$, (corresponding Rabi constant and plasmon decay rate are $\OR =0.21\cdot 10^{13} s^{-1} $ and $\gamma _{\rm pl} =0.46\cdot 10^{13} s^{-1} $, respectively).\label{Fig3}}
\end{figure*}
\begin{figure*}[t]
\includegraphics[width=\textwidth]{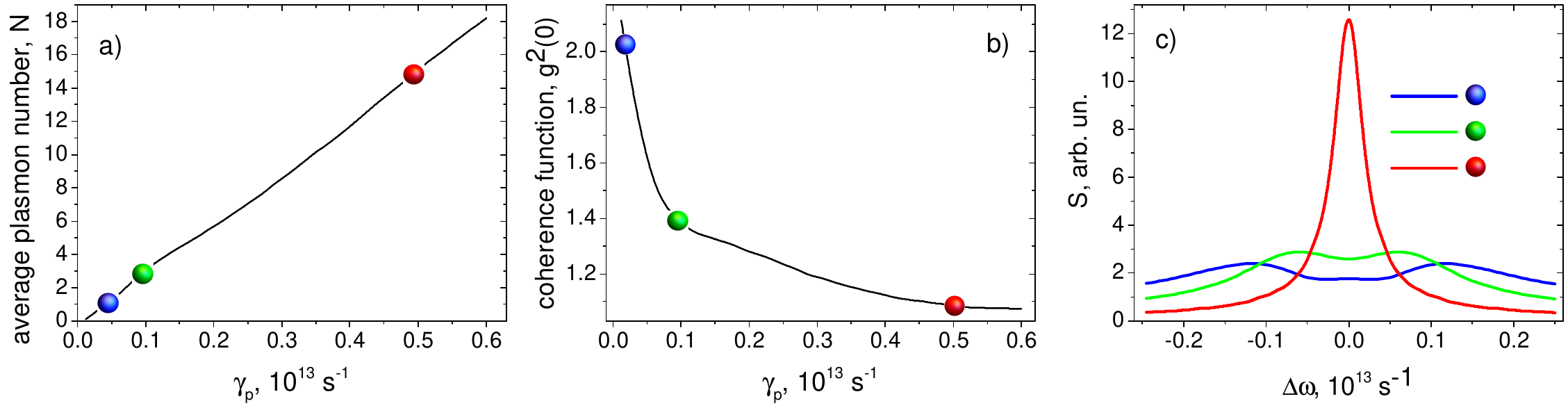}
\caption{The same as in Fig.~\ref{Fig3}. Parameters of graphene are the following: $\Ef \simeq 0.5eV$, $\lambda _{\rm pl} =4\mum$, (corresponding Rabi constant and plasmon decay rate are $\OR =0.43\cdot 10^{13} s^{-1} $ and $\gamma _{\rm pl} =0.64\cdot 10^{13} s^{-1} $, respectively).\label{Fig4}}
\end{figure*}
\begin{figure*}[t]
\includegraphics[width=\textwidth]{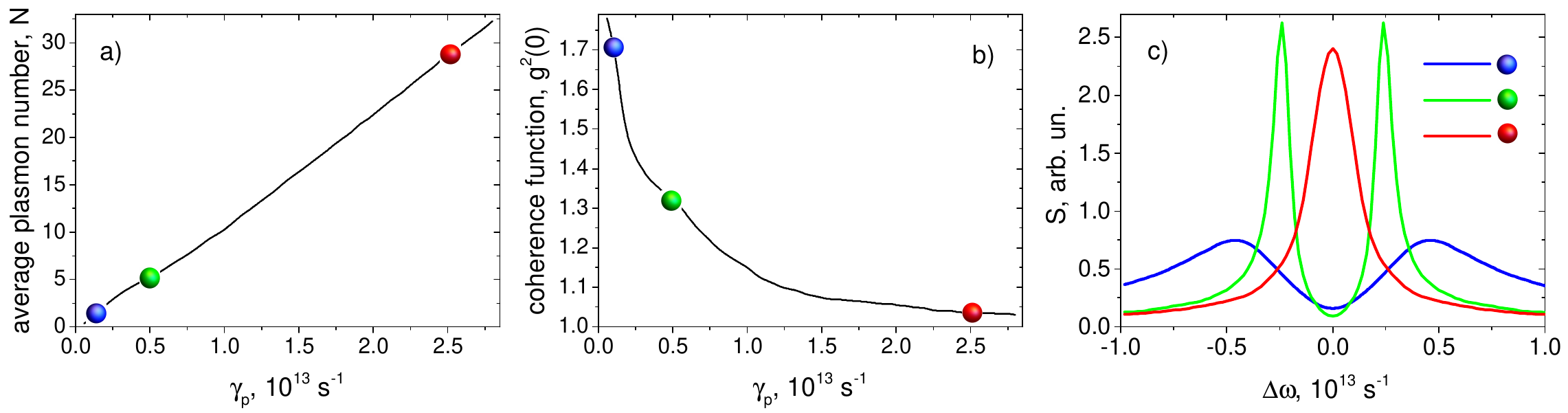}
\caption{(Color online) The same as in Fig.~\ref{Fig3}. Parameters of graphene are the following: $\Ef \simeq 0.4eV$, $\lambda _{\rm pl} =3\mum$, (corresponding Rabi constant and plasmon decay rate are $\OR =1.56\cdot 10^{13} s^{-1} $ and $\gamma _{\rm pl} =1.76\cdot 10^{13} s^{-1} $, respectively).\label{Fig5}}
\end{figure*}

To calculate interaction constant between plasmon field and graphene we use standard relation
\begin{gather} \label{30}
	\OR=  - {{\bf{d}}_{12}}{\bf{E}}/\hbar,
\end{gather}
where ${{\bf{d}}_{12}}$ is matrix element of dipole moment, {\bf{E}} is the field amplitude. It can be determined from the Helmholtz equation
\begin{gather} \label{301}
\nabla  \times \nabla  \times {\bf{E}} - \frac{{\omega_{\rm pl}^2}}{{{c^2}}}{\varepsilon }\left( {\bf{r}} \right){\bf{E}} = 0,
\end{gather}
and normalized through
\begin{gather} \label{31}
	\frac{1}{{8\pi }}\int {\left[ {\frac{{\partial \left( {\varepsilon '\omega } \right)}}{{\partial \omega }}\left( {{\bf{E}}{{\bf{E}}^ * }} \right) + \left( {{\bf{H}}{{\bf{H}}^ * }} \right)} \right]dV}  = \hbar {\omega_{\rm pl}}.
\end{gather}
Note that
\begin{multline} \label{32}
\frac{1}{{8\pi }}\int {\left[ {\frac{{\partial \left( {\varepsilon'\omega } \right)}}{{\partial \omega }}\left( {{\bf{E}}{{\bf{E}}^* }} \right) + \left( {{\bf{H}}{{\bf{H}}^* }} \right)} \right]dV} =
\\
\frac{1}{{8\pi {\omega_{\rm pl}}}}\int {{{\left. {\frac{{\partial \left( {\varepsilon'{\omega^2}} \right)}}{{\partial \omega }}} \right|}_{{\omega_{\rm pl}}}}\left( {{\bf{E}}{{\bf{E}}^* }} \right)dV}.
\end{multline}

In the structure under study, see Fig.~\ref{fignanolaser}, the permittivity $\varepsilon  = \varepsilon \left( {\omega ,z} \right)$  essentially depends on coordinate. At $z < 0$ it  is equal to the
permittivity of the substrate, at $z = 0$ it is equal to the surface permittivity of graphene: $\varepsilon  = \delta \left( z \right)4\pi i\sigma /\omega $, where $\delta \left( z \right)$  is the Dirac delta function and $\sigma $  is the graphene surface conductivity; at $z > 0$ the permittivity is equal to unity. Finally we have
\begin{multline}
\int {{{\left. {\frac{{\partial \left( {\varepsilon '{\omega^2}} \right)}}{{\partial \omega }}} \right|}_{{\omega_{\rm pl}}}}\left( {{\bf{E}}{{\bf{E}}^ * }} \right)dV}  =- 4\pi {\left. {\frac{{\partial \left( {\sigma ''\omega } \right)}}{{\partial \omega }}} \right|_{{\omega_{\mathrm{pl}}}}}\left| {{E_t}} \right|_g^2A+
\\
2A\omega \int\limits_0^\infty  {\left( {{\bf{E}}{{\bf{E}}^* }} \right)dz}  + 2A\omega {\varepsilon'_{\scriptscriptstyle \mathrm D}}\int\limits_{ - \infty }^0 {\left( {{\bf{E}}{{\bf{E}}^ * }} \right)dz},
\end{multline}
where $\epsD$  is permittivity of the dielectric substrate, $A$ is the area of graphene nanoflake, $E_t$ is the tangential component of electric field, subscript $g$ denotes the value at $z=0$ on graphene layer.

Let $\kappa $  and ${\kappaD}$ are the imaginary parts of the normal components of surface plasmon wave vectors in vacuum and dielectric correspondingly. So ${\kappa^2} = {k^2} - k_0^2$  and $\kappaD^2 = {k^2} - k_0^2\epsD$ , where $k$ is the surface plasmon wave number and ${k_0} = \omega /c$ .

 Electric field depends on $z$ as follows: $\left( {{\bf{E}}{{\bf{E}}^ * }} \right) = \left| {\bf{E}} \right|_g^2\exp \left( {2{\kappaD}z} \right)$  in dielectric ($z < 0$) and $\left( {{\bf{E}}{{\bf{E}}^ * }} \right) = \left| {\bf{E}} \right|_g^2\exp \left( { - 2\kappa z} \right)$ in vacuum ($z>0$). Then
\begin{multline}
\int {{{\left. {\frac{{\partial \left( {\varepsilon '{\omega ^2}} \right)}}{{\partial \omega }}} \right|}_{{\omega_0}}}\left( {{\bf{E}}{{\bf{E}}^* }} \right)dV}  = - 4\pi {\left. {\frac{{\partial \left( {\sigma''\omega } \right)}}{{\partial \omega }}} \right|_{{\omega_0}}}\left| {{E_t}} \right|_g^2A +
\\
 2A\omega \left| {\bf{E}} \right|_g^2/\left( {2\kappa } \right) + 2A\omega {\varepsilon'_{\mathrm{\scriptscriptstyle D}}}\left| {\bf{E}} \right|_g^2/\left( {2{\kappaD}} \right).
\end{multline}
Finally, let us note the relation between the tangential and normal electric field components: $k{E_t} + i\kappa {E_n} = 0$  in vacuum and $k{E_t} + i{\kappaD}{E_n} = 0$  in dielectric. Along with the relation ${\left| {\bf{E}} \right|^2} = {\left| {{E_t}} \right|^2} + {\left| {{E_n}} \right|^2}$ , we find $\left| {\bf{E}} \right|_g^2 = \left| {{E_t}} \right|_g^2\frac{{2{k^2} - k_0^2}}{{{k^2} - k_0^2}}$  in vacuum and $\left| {\bf{E}} \right|_g^2 = \left| {{E_t}} \right|_g^2\frac{{2{k^2} - k_0^2\epsD}}{{{k^2} - k_0^2\epsD}}$ in dielectric. For plasmon in graphene, ${k^2} \gg k_0^2$, which means $\left| {\bf{E}} \right|_g^2 \approx 2\left| {{E_t}} \right|_g^2$ . In the same approximation, ${\kappa ^2} \approx \kappaD^2 \approx {k^2}$ , which leads to
\begin{multline}
\int {{{\left. {\frac{{\partial \left( {\varepsilon'{\omega^2}} \right)}}{{\partial \omega }}} \right|}_{{\omega_{\rm pl}}}}\left( {{\bf{E}}{{\bf{E}}^ * }} \right)dV}  =
\\
\left[ { - 4\pi {{\left. {\frac{{\partial \left( {\sigma''\omega } \right)}}{{\partial \omega }}} \right|}_{{\omega_{\rm pl}}}} + 4\omega /k} \right]\left| {{E_t}} \right|_g^2A.
\end{multline}
 Using this equation and relation \eqref{31} we transform \eqref{30} into
\begin{gather}\label{ApOmegaR}
	\OR =  - {d_{12}}{\omega_{\rm pl}}\sqrt {\frac{{2k^3 }}{{\pi\hbar \left[ {- \pi k \left.\frac{\partial }{\partial \omega} (\sigma''\omega)\right|_{\omega_{\rm pl}} + \omega_{\rm pl}} \right]}}}\,,
\end{gather}
where we take into account that $A=(\lambda/2)^2 = (\pi/k\omega_{\rm pl})^2$ to satisfy resonant condition. Conductivity for graphene layer may be expressed as~\cite{falkovsky2008PhysUsp,kotov2013OptExp}:
\begin{align}\label{Apsigma}
\sigma& \left( {\omega, {\Ef}} \right)/{e^2}=\frac12 +\frac1\pi \arctan\left [\frac{\hbar \omega  -2{\Ef}}{2{\kB}T}\right] +
\\\notag
&\qquad\frac i{\pi}\left\{\frac{ 8{\kB}T\ln\left [2\cosh\left(\frac{{\Ef}}{2{\kB}T}\right)\right]}{\hbar \left(\omega  + \frac i\tau \right)}+\right.
\\\notag
 &\qquad\qquad\log\left [\frac{\left(\hbar (\omega  +\frac i\tau ) + 2\Ef\right)^2}{\left(\hbar (\omega  +\frac i\tau ) + 2\Ef\right)^2+(2\kB T)^2}\right]\Biggl\}.
\end{align}
So from \eqref{ApOmegaR} and \eqref{Apsigma} we obtain the dependence of Rabi constant on the plasmon frequency and Fermi energy, see Fig.~\ref{gammaPlOmegaR}.

\subsection{ Plasmon decay rate $\gamma_{\rm pl} $}
The plasmonic mode relaxation rate due to Joule losses is introduced as~\cite{landau1984electrodynamics,novotny2012principles}
\begin{gather}
{\gamma_{\rm pl}} = \frac{{{\omega_{\rm pl}}}}{{4\pi }}\int {{\varepsilon}^{\prime \prime }\left( {{\bf{E}}{{\bf{E}}^* }} \right)dV} /W
\end{gather}
Using the relations \eqref{31} and \eqref{32} and taking into account that $\varepsilon'' = \delta \left( z \right)4\pi \sigma'/\omega $ we obtain
\begin{gather}
	\gamma_{\rm pl}  = \frac{2\pi k\omega_{\rm pl}\sigma'}{- \pi k \left.\frac{\partial }{\partial \omega} (\sigma''\omega)\right|_{\omega_{\rm pl}} + 0.5\omega_{\rm pl}}.
\end{gather}
So, we have found the dependence of Rabi constant of interaction between plasmon field and active medium, and plasmon relaxation rate as function of $\lambda_{pl}$ and $E_F$.

\section{Transition from enhanced spontaneous emission to lasing}

Now we investigate the generation of surface plasmon and its coherent properties using parameters determined in the previous section. First we should note that all the equations derived above are applicable when the Fermi level is in the  range $\sim 0.1-1\, eV$ and plasmon wavelength $\sim 1-10\, \mum$: otherwise we should take into account intraband transitions in graphene stimulated pumping.
Doing numerical calculation we see that the Rabi constant and the damping rate have qualitatively the same behavior: they decrease as the functions of $E_F$ and $\lambda_{pl}$, see Fig.~\ref{gammaPlOmegaR}.

Now we will investigate the graphene nanolaser dynamics. For this we solve master equation \eqref{GrindEQ__1_} changing parameters $\gamma _{\rm pl} $ and $\OR $.

Sometimes it is a difficult question how to distinguish lasing from highly coherent fluorescence. There are several important observables describing the transition to lasing behavior~\cite{haken1983laser,siegman1986lasers,khanin2005fundamentals,chow2014LSA}. The first one is the dependence of the plasmon number $N_{\rm pl} =\left\langle \hat{a}^{+} \hat{a}\right\rangle =\Tr\left(\hat{\rho }\hat{a}^{+} \hat{a}\right)$ on the pumping rate $\gamma_{\rm am}^{\rm pump} $ in double logarithmic scale. ``S'' shape of such curve indicates the first lasing threshold. (There are in fact threshholdless lasers with no pronounced lower threshold~\cite{protsenko2012theory,protsenko1999PRA,protsenko2005PRA}.) The second characteristic is the dependence of the second order correlation function, $g^{(2)} (0)=\left\langle \hat{a}^{+} \hat{a}^{+} \hat{a}\hat{a}\right\rangle /\left\langle \hat{a}^{+} \hat{a}\right\rangle ^{2} $. When $g^{(2)} (0)=1$ we have Poisson photon statistics; usually in lasing systems this case corresponds to coherent photon state. If $g^{(2)} (0)>1$ then the photon state is ``incoherent'' and statistics is super-Poisson. If $g^{(2)} (0)<1$ then statistics is sub-Poison.

One more characteristics which we will use is spectrum of the generated plasmon, namely, $S(\omega )=\mathrm{Re}\int _{0}^{\infty }d\tau \left\langle \hat{a}^{+} (t+\tau )\hat{a}(t)\right\rangle \exp \left(i\omega \tau \right) $. Narrowing of the spectral line may be interpreted as increasing of coherence. The value of pumping rate corresponding to narrowing of the spectral line width we refer to as the second threshold.

\begin{figure}[t]
\includegraphics*[width=\columnwidth]{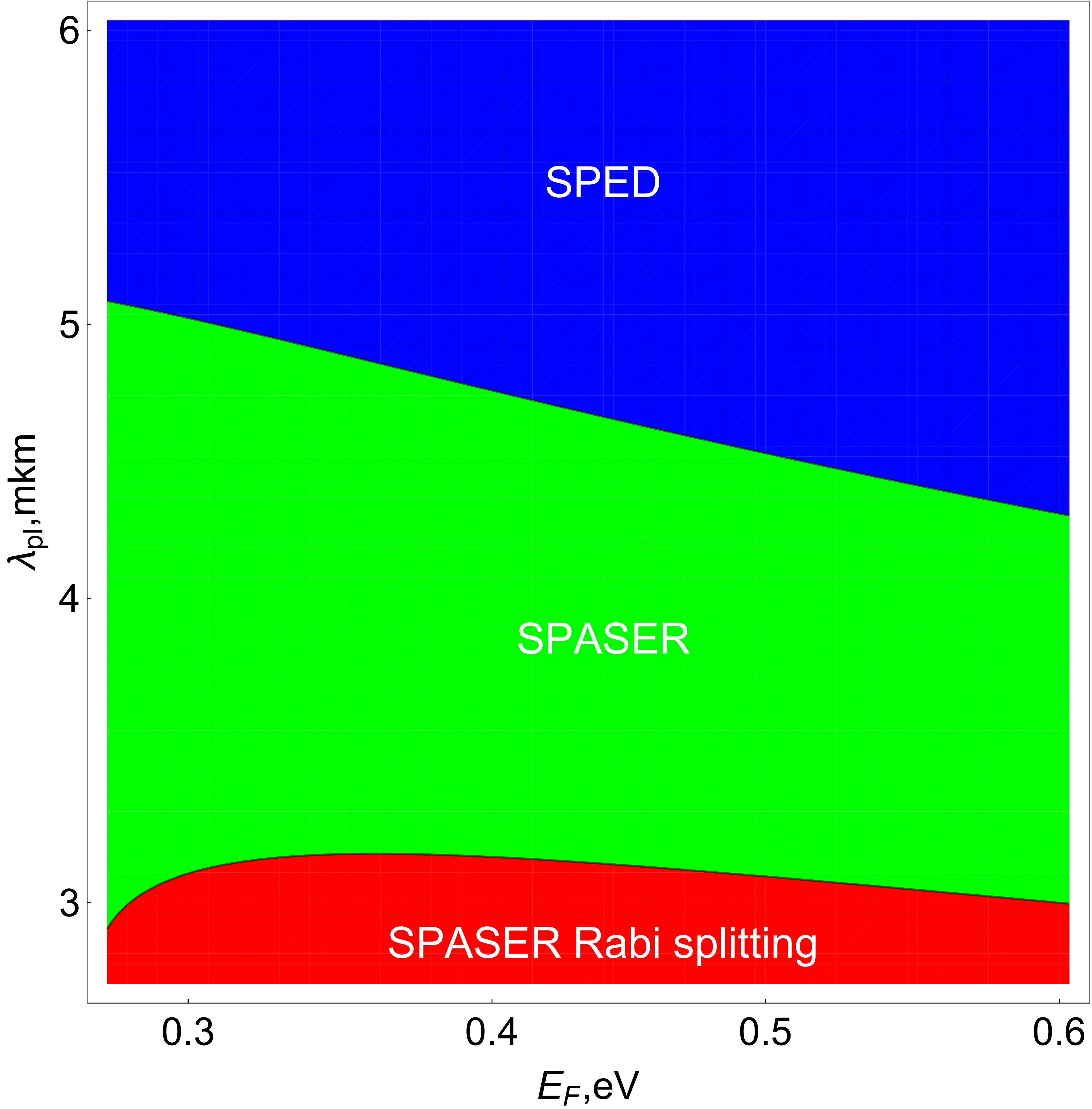}
\caption{(Color online) The dependence of the plasmon decay rate $\gamma _{\rm pl} $ and Rabi constant $\OR $ in the dephasing rate $\gamma _{\rm am}^{\rm dephasing} $ units. Vertical lines distinguish three working regimes of graphene nanolaser.\label{Fig6}}
\end{figure}
Doing numerical calculations we assume that the number of two-level particles of active medium which interact with plasmonic mode is $N = 20$. This is so because the characteristic length of the graphene sheet $\sim 1\mu m $ and the size of usual mid-IR quantum dots is about $\sim 20$nm and distance between them is the same order.

There are three different regimes. The first one corresponds to the high plasmon wavelength, namely, $5\mum<\lambda_{\rm pl}$. The second regime corresponds to $3\mum<\lambda_{\rm pl} <5$$\mu m$. In the third case we have $\lambda_{\rm pl} <3\mum$.

In the first regime, see Fig.~\ref{Fig3}, we have thresholdless behavior of the plasmon number on pumping, whereas the $g^{(2)} (0)$ does not approach the unity. At the same time, line narrowing is not pronounced. In this case we have  enhanced spontaneous excitation of surface plasmon rather then stimulated emission of them. Device based on noncoherent generation of surface plasmons may be named as surface plasmon emitting diode ~\cite{khurgin2014NatPhot} in analogous with light emitting diode in optical case.

At second stage, see Fig.~\ref{Fig4}, we obtain thresholdless behavior of the plasmon number which indicates that there is relatively large number of plasmons excited by spontaneous transitions in active medium. So the dependence of the plasmon number on pumping does not indicate any coherence. However, other characteristics exhibit coherence explicitly: $g^{(2)} (0)$ tends to unity and spectral line strongly narrows at the same time. These characteristics indicate the possibility of generation of coherent plasmons. So, in this case we have SPASER - surface plasmon generator by stimulated emission of radiation.

Third regime, see Fig.~\ref{Fig5}, is very similar to the previous one: $g^{(2)} (0)$ tends to unity. However, line narrowing demonstrates one interesting feature. At moderate values of pumping we have doublet structure with two narrow lines. This happens due to high value of Rabi frequency at low value of surface plasmon wavelength. We name this regime as SPASER with Rabi spliting.

In Fig.~\ref{Fig6} we summarize the main results. So, we consider three different regimes: SPED (surface plasmon emitting diode), when there is high value of spontaneous emission, SPASER (surface plasmon amplification by stimulated emission of radiation) regime, at which there is possible to obtain coherence at realistic pump rates and SPASER with Rabi splitting  which is characterized by two narrow line of spectrum.

Note that for each plasmon wavelength it is necessary to use appropriate active medium with suitable transition frequency between working levels. Good candidates for this purpose are various types of quantum dots. For example, for SPASER and SPASER Rabi splitting regime PbSe ~\cite{pietryga2004pushing,wehrenberg2002JCP} and HgTe colloidal quantum dots ~\cite{keuleyan2011synthesis,lhuiller2013ChemMat} as well as $\mathrm{Cr}^{2+}$ dopped ZnS, ZnSe, CdSe ~\cite{mirov2010LPR}, $\mathrm{Fe}^{2+}$ dopped ZnSe, CdMnTe ~\cite{mirov2010LPR}, SiGe quantum dots ~\cite{wang2007IEEE}, InGaAs/GaAs quantum box structure ~\cite{botez2007INJ} and transition-metal-doped nanocrystalline quantum dots ~\cite{mirov2007JSTQE} may be used. For SPED regime HgTe colloidal quantum dots and transition-metal-doped nanocrystalline quantum dots (QD) is also appropriate. We summarize these data in Table 1.

\begin{table}
	\centering\
	\caption{Active mediums and working wavelength}
	\begin{tabular}{|c|c|}
	\hline $\lambda$, $\mum$ & Type of active medium \\
	\hline 1.5 - 5 & HgTe colloidal QDs ~\cite{keuleyan2011synthesis,lhuiller2013ChemMat} \\
	\hline 2 - 4 & PbSe colloidal QDs ~\cite{pietryga2004pushing,wehrenberg2002JCP}  \\
	\hline 2.5 & $\mathrm{Cr}^{2+}$ dopped ZnS, ZnSe, CdSe ~\cite{mirov2010LPR} \\
	\hline 3 - 5& Transition-metal-doped nanocrystalline QDs ~\cite{mirov2007JSTQE}   \\
	\hline 3.5 & SiGe QDs ~\cite{wang2007IEEE} \\
	\hline 4.5 & $\mathrm{Fe}^{2+}$ dopped ZnSe, CdMnTe ~\cite{mirov2010LPR} \\
	\hline  4.7 & InGaAs/GaAs quantum box structure ~\cite{botez2007INJ} \\
	\hline
	\end{tabular}
\end{table}

\section{Conclusions}

In conclusion, we have shown doing self-consistent quantum calculations that graphene is the promising material for applications in state-of-the-art active and passive plasmonic devices that allow \textit{in situ} tuning of parameters. High graphene conductivity dependence on Fermi-level and frequency allows switching between such generation types such as SPED and SPASER (surface plasmon amplification by stimulated emission of radiation). Corresponding generation spectrum and the second order correlation function which predicts laser statistics have been calculated. We provide explicit expressions for interaction and dissipation parameters through material constants and geometry.

\acknowledgments
The work was partially funded by Russian Foundation for Scientific Research (grant No.16-02-00295), while numerical simulations were funded by Russian Scientific Foundation (grant No.14-12-01185). We thank Supercomputer Centers of Russian Academy of Sciences and National research Center Kurchatov Institute for access to URAL, JSCC and HPC supercomputer clusters.

\noindent
\bibliography{refs}

\begin{thebibliography}{88}%
\makeatletter
\providecommand \@ifxundefined [1]{%
 \@ifx{#1\undefined}
}%
\providecommand \@ifnum [1]{%
 \ifnum #1\expandafter \@firstoftwo
 \else \expandafter \@secondoftwo
 \fi
}%
\providecommand \@ifx [1]{%
 \ifx #1\expandafter \@firstoftwo
 \else \expandafter \@secondoftwo
 \fi
}%
\providecommand \natexlab [1]{#1}%
\providecommand \enquote  [1]{``#1''}%
\providecommand \bibnamefont  [1]{#1}%
\providecommand \bibfnamefont [1]{#1}%
\providecommand \citenamefont [1]{#1}%
\providecommand \href@noop [0]{\@secondoftwo}%
\providecommand \href [0]{\begingroup \@sanitize@url \@href}%
\providecommand \@href[1]{\@@startlink{#1}\@@href}%
\providecommand \@@href[1]{\endgroup#1\@@endlink}%
\providecommand \@sanitize@url [0]{\catcode `\\12\catcode `\$12\catcode
  `\&12\catcode `\#12\catcode `\^12\catcode `\_12\catcode `\%12\relax}%
\providecommand \@@startlink[1]{}%
\providecommand \@@endlink[0]{}%
\providecommand \url  [0]{\begingroup\@sanitize@url \@url }%
\providecommand \@url [1]{\endgroup\@href {#1}{\urlprefix }}%
\providecommand \urlprefix  [0]{URL }%
\providecommand \Eprint [0]{\href }%
\providecommand \doibase [0]{http://dx.doi.org/}%
\providecommand \selectlanguage [0]{\@gobble}%
\providecommand \bibinfo  [0]{\@secondoftwo}%
\providecommand \bibfield  [0]{\@secondoftwo}%
\providecommand \translation [1]{[#1]}%
\providecommand \BibitemOpen [0]{}%
\providecommand \bibitemStop [0]{}%
\providecommand \bibitemNoStop [0]{.\EOS\space}%
\providecommand \EOS [0]{\spacefactor3000\relax}%
\providecommand \BibitemShut  [1]{\csname bibitem#1\endcsname}%
\let\auto@bib@innerbib\@empty
%</preamble>
\bibitem [{\citenamefont {Maier}(2007)}]{maier2007plasmonics}%
  \BibitemOpen
  \bibfield  {author} {\bibinfo {author} {\bibfnamefont {S.~A.}\ \bibnamefont
  {Maier}},\ }\href@noop {} {\emph {\bibinfo {title} {Plasmonics: fundamentals
  and applications}}}\ (\bibinfo  {publisher} {Springer Science \& Business
  Media},\ \bibinfo {year} {2007})\BibitemShut {NoStop}%
\bibitem [{\citenamefont {Shvets}\ and\ \citenamefont
  {Tsukerman}(2011)}]{shvets2011plasmonics}%
  \BibitemOpen
  \bibfield  {author} {\bibinfo {author} {\bibfnamefont {G.}~\bibnamefont
  {Shvets}}\ and\ \bibinfo {author} {\bibfnamefont {I.}~\bibnamefont
  {Tsukerman}},\ }\href@noop {} {\emph {\bibinfo {title} {Plasmonics and
  Plasmonic Metamaterials: Analysis and Applications}}},\ Vol.~\bibinfo
  {volume} {4}\ (\bibinfo  {publisher} {World Scientific},\ \bibinfo {year}
  {2011})\BibitemShut {NoStop}%
\bibitem [{\citenamefont {Enoch}\ and\ \citenamefont
  {Bonod}(2012)}]{enoch2012plasmonics}%
  \BibitemOpen
  \bibfield  {author} {\bibinfo {author} {\bibfnamefont {S.}~\bibnamefont
  {Enoch}}\ and\ \bibinfo {author} {\bibfnamefont {N.}~\bibnamefont {Bonod}},\
  }\href@noop {} {\emph {\bibinfo {title} {Plasmonics: from basics to advanced
  topics}}},\ Vol.\ \bibinfo {volume} {167}\ (\bibinfo  {publisher}
  {Springer},\ \bibinfo {year} {2012})\BibitemShut {NoStop}%
\bibitem [{\citenamefont {Zouhdi}\ \emph {et~al.}(2008)\citenamefont {Zouhdi},
  \citenamefont {Sihvola},\ and\ \citenamefont
  {Vinogradov}}]{zouhdi2008metamaterials}%
  \BibitemOpen
  \bibfield  {author} {\bibinfo {author} {\bibfnamefont {S.}~\bibnamefont
  {Zouhdi}}, \bibinfo {author} {\bibfnamefont {A.}~\bibnamefont {Sihvola}}, \
  and\ \bibinfo {author} {\bibfnamefont {A.~P.}\ \bibnamefont {Vinogradov}},\
  }\href@noop {} {\emph {\bibinfo {title} {Metamaterials and Plasmonics:
  Fundamentals, Modelling, Applications}}}\ (\bibinfo  {publisher} {Springer
  Science \& Business Media},\ \bibinfo {year} {2008})\BibitemShut {NoStop}%
\bibitem [{\citenamefont {Bozhevolnyi}(2006)}]{bozhevolnyiplasmonic}%
  \BibitemOpen
  \bibfield  {author} {\bibinfo {author} {\bibfnamefont {S.}~\bibnamefont
  {Bozhevolnyi}},\ }\href@noop {} {\enquote {\bibinfo {title} {Plasmonic
  nanoguides and circuits. 2008},}\ } (\bibinfo {year} {2006})\BibitemShut
  {NoStop}%
\bibitem [{\citenamefont {Shalaev}\ and\ \citenamefont
  {Kawata}(2006)}]{shalaev2006nanophotonics}%
  \BibitemOpen
  \bibfield  {author} {\bibinfo {author} {\bibfnamefont {V.~M.}\ \bibnamefont
  {Shalaev}}\ and\ \bibinfo {author} {\bibfnamefont {S.}~\bibnamefont
  {Kawata}},\ }\href@noop {} {\emph {\bibinfo {title} {Nanophotonics with
  surface plasmons}}}\ (\bibinfo  {publisher} {Elsevier},\ \bibinfo {year}
  {2006})\BibitemShut {NoStop}%
\bibitem [{\citenamefont {Zayats}\ and\ \citenamefont
  {Maier}(2013)}]{zayats2013active}%
  \BibitemOpen
  \bibfield  {author} {\bibinfo {author} {\bibfnamefont {A.~V.}\ \bibnamefont
  {Zayats}}\ and\ \bibinfo {author} {\bibfnamefont {S.}~\bibnamefont {Maier}},\
  }\href@noop {} {\emph {\bibinfo {title} {Active plasmonics and tuneable
  plasmonic metamaterials}}},\ Vol.~\bibinfo {volume} {8}\ (\bibinfo
  {publisher} {John Wiley \& Sons},\ \bibinfo {year} {2013})\BibitemShut
  {NoStop}%
\bibitem [{\citenamefont {Sarychev}\ and\ \citenamefont
  {Shalaev}(2007)}]{sarychev2007electrodynamics}%
  \BibitemOpen
  \bibfield  {author} {\bibinfo {author} {\bibfnamefont {A.~K.}\ \bibnamefont
  {Sarychev}}\ and\ \bibinfo {author} {\bibfnamefont {V.~M.}\ \bibnamefont
  {Shalaev}},\ }\href@noop {} {\emph {\bibinfo {title} {Electrodynamics of
  metamaterials}}}\ (\bibinfo  {publisher} {World Scientific},\ \bibinfo {year}
  {2007})\BibitemShut {NoStop}%
\bibitem [{\citenamefont {West}\ \emph {et~al.}(2010)\citenamefont {West},
  \citenamefont {Ishii}, \citenamefont {Naik}, \citenamefont {Emani},
  \citenamefont {Shalaev},\ and\ \citenamefont {Boltasseva}}]{west2010LPR}%
  \BibitemOpen
  \bibfield  {author} {\bibinfo {author} {\bibfnamefont {P.~R.}\ \bibnamefont
  {West}}, \bibinfo {author} {\bibfnamefont {S.}~\bibnamefont {Ishii}},
  \bibinfo {author} {\bibfnamefont {G.~V.}\ \bibnamefont {Naik}}, \bibinfo
  {author} {\bibfnamefont {N.~K.}\ \bibnamefont {Emani}}, \bibinfo {author}
  {\bibfnamefont {V.~M.}\ \bibnamefont {Shalaev}}, \ and\ \bibinfo {author}
  {\bibfnamefont {A.}~\bibnamefont {Boltasseva}},\ }\href@noop {} {\bibfield
  {journal} {\bibinfo  {journal} {Laser \& Photonics Reviews}\ }\textbf
  {\bibinfo {volume} {4}},\ \bibinfo {pages} {795} (\bibinfo {year}
  {2010})}\BibitemShut {NoStop}%
\bibitem [{\citenamefont {E.}\ and\ \citenamefont {V.}(1987)}]{Lozovik1987}%
  \BibitemOpen
  \bibfield  {author} {\bibinfo {author} {\bibfnamefont {L.~Y.}\ \bibnamefont
  {E.}}\ and\ \bibinfo {author} {\bibfnamefont {K.~A.}\ \bibnamefont {V.}},\
  }\href@noop {} {\emph {\bibinfo {title} {The dielectric function and
  collective oscillations inhomogeneous systems}}}\ (\bibinfo  {publisher} {ed.
  by Keldysh, L. V. and Kirzhnits D. A. and Maradudin A. A., North Holland:
  Amsterdam},\ \bibinfo {year} {1987})\BibitemShut {NoStop}%
\bibitem [{\citenamefont {Ji}\ \emph {et~al.}(2007)\citenamefont {Ji},
  \citenamefont {Xie},\ and\ \citenamefont {Liu}}]{Liu2007PhysRevLett}%
  \BibitemOpen
  \bibfield  {author} {\bibinfo {author} {\bibfnamefont {A.-C.}\ \bibnamefont
  {Ji}}, \bibinfo {author} {\bibfnamefont {X.~C.}\ \bibnamefont {Xie}}, \ and\
  \bibinfo {author} {\bibfnamefont {W.~M.}\ \bibnamefont {Liu}},\ }\href
  {\doibase 10.1103/PhysRevLett.99.183602} {\bibfield  {journal} {\bibinfo
  {journal} {Phys. Rev. Lett.}\ }\textbf {\bibinfo {volume} {99}},\ \bibinfo
  {pages} {183602} (\bibinfo {year} {2007})}\BibitemShut {NoStop}%
\bibitem [{\citenamefont {Barnes}\ \emph {et~al.}(2003)\citenamefont {Barnes},
  \citenamefont {Dereux},\ and\ \citenamefont {Ebbesen}}]{barnes2003Nature}%
  \BibitemOpen
  \bibfield  {author} {\bibinfo {author} {\bibfnamefont {W.~L.}\ \bibnamefont
  {Barnes}}, \bibinfo {author} {\bibfnamefont {A.}~\bibnamefont {Dereux}}, \
  and\ \bibinfo {author} {\bibfnamefont {T.~W.}\ \bibnamefont {Ebbesen}},\
  }\href@noop {} {\bibfield  {journal} {\bibinfo  {journal} {Nature}\ }\textbf
  {\bibinfo {volume} {424}},\ \bibinfo {pages} {824} (\bibinfo {year}
  {2003})}\BibitemShut {NoStop}%
\bibitem [{\citenamefont {Radko}\ \emph {et~al.}(2008)\citenamefont {Radko},
  \citenamefont {Evlyukhin}, \citenamefont {Boltasseva},\ and\ \citenamefont
  {Bozhevolnyi}}]{radko2008OptExp}%
  \BibitemOpen
  \bibfield  {author} {\bibinfo {author} {\bibfnamefont {I.~P.}\ \bibnamefont
  {Radko}}, \bibinfo {author} {\bibfnamefont {A.~B.}\ \bibnamefont
  {Evlyukhin}}, \bibinfo {author} {\bibfnamefont {A.}~\bibnamefont
  {Boltasseva}}, \ and\ \bibinfo {author} {\bibfnamefont {S.~I.}\ \bibnamefont
  {Bozhevolnyi}},\ }\href@noop {} {\bibfield  {journal} {\bibinfo  {journal}
  {Opt. Exp.}\ }\textbf {\bibinfo {volume} {16}},\ \bibinfo {pages} {3924}
  (\bibinfo {year} {2008})}\BibitemShut {NoStop}%
\bibitem [{\citenamefont {Gong}\ and\ \citenamefont
  {Vu{\v{c}}kovi{\'c}}(2007)}]{gong2007APL}%
  \BibitemOpen
  \bibfield  {author} {\bibinfo {author} {\bibfnamefont {Y.}~\bibnamefont
  {Gong}}\ and\ \bibinfo {author} {\bibfnamefont {J.}~\bibnamefont
  {Vu{\v{c}}kovi{\'c}}},\ }\href@noop {} {\bibfield  {journal} {\bibinfo
  {journal} {Appl. Phys. Lett.}\ }\textbf {\bibinfo {volume} {90}},\ \bibinfo
  {pages} {033113} (\bibinfo {year} {2007})}\BibitemShut {NoStop}%
\bibitem [{\citenamefont {Archambault}\ \emph {et~al.}(2009)\citenamefont
  {Archambault}, \citenamefont {Teperik}, \citenamefont {Marquier},\ and\
  \citenamefont {Greffet}}]{archambault2009PRB}%
  \BibitemOpen
  \bibfield  {author} {\bibinfo {author} {\bibfnamefont {A.}~\bibnamefont
  {Archambault}}, \bibinfo {author} {\bibfnamefont {T.~V.}\ \bibnamefont
  {Teperik}}, \bibinfo {author} {\bibfnamefont {F.}~\bibnamefont {Marquier}}, \
  and\ \bibinfo {author} {\bibfnamefont {J.-J.}\ \bibnamefont {Greffet}},\
  }\href@noop {} {\bibfield  {journal} {\bibinfo  {journal} {Phys. Rev. B}\
  }\textbf {\bibinfo {volume} {79}},\ \bibinfo {pages} {195414} (\bibinfo
  {year} {2009})}\BibitemShut {NoStop}%
\bibitem [{\citenamefont {Feng}\ \emph {et~al.}(2007)\citenamefont {Feng},
  \citenamefont {Tetz}, \citenamefont {Slutsky}, \citenamefont {Lomakin},\ and\
  \citenamefont {Fainman}}]{feng2007APL}%
  \BibitemOpen
  \bibfield  {author} {\bibinfo {author} {\bibfnamefont {L.}~\bibnamefont
  {Feng}}, \bibinfo {author} {\bibfnamefont {K.~A.}\ \bibnamefont {Tetz}},
  \bibinfo {author} {\bibfnamefont {B.}~\bibnamefont {Slutsky}}, \bibinfo
  {author} {\bibfnamefont {V.}~\bibnamefont {Lomakin}}, \ and\ \bibinfo
  {author} {\bibfnamefont {Y.}~\bibnamefont {Fainman}},\ }\href@noop {}
  {\bibfield  {journal} {\bibinfo  {journal} {Appl. Phys. Lett.}\ }\textbf
  {\bibinfo {volume} {91}},\ \bibinfo {pages} {081101} (\bibinfo {year}
  {2007})}\BibitemShut {NoStop}%
\bibitem [{\citenamefont {Zia}\ and\ \citenamefont
  {Brongersma}(2007)}]{zia2007NatNano}%
  \BibitemOpen
  \bibfield  {author} {\bibinfo {author} {\bibfnamefont {R.}~\bibnamefont
  {Zia}}\ and\ \bibinfo {author} {\bibfnamefont {M.~L.}\ \bibnamefont
  {Brongersma}},\ }\href@noop {} {\bibfield  {journal} {\bibinfo  {journal}
  {Nat. Nanotechnology}\ }\textbf {\bibinfo {volume} {2}},\ \bibinfo {pages}
  {426} (\bibinfo {year} {2007})}\BibitemShut {NoStop}%
\bibitem [{\citenamefont {Pockrand}\ \emph {et~al.}(1978)\citenamefont
  {Pockrand}, \citenamefont {Swalen}, \citenamefont {Gordon},\ and\
  \citenamefont {Philpott}}]{pockrand1978surface}%
  \BibitemOpen
  \bibfield  {author} {\bibinfo {author} {\bibfnamefont {I.}~\bibnamefont
  {Pockrand}}, \bibinfo {author} {\bibfnamefont {J.}~\bibnamefont {Swalen}},
  \bibinfo {author} {\bibfnamefont {J.}~\bibnamefont {Gordon}}, \ and\ \bibinfo
  {author} {\bibfnamefont {M.}~\bibnamefont {Philpott}},\ }\href@noop {}
  {\bibfield  {journal} {\bibinfo  {journal} {Surface Science}\ }\textbf
  {\bibinfo {volume} {74}},\ \bibinfo {pages} {237} (\bibinfo {year}
  {1978})}\BibitemShut {NoStop}%
\bibitem [{\citenamefont {Mills}\ and\ \citenamefont
  {Agranovich}(1982)}]{mills1982surface}%
  \BibitemOpen
  \bibfield  {author} {\bibinfo {author} {\bibfnamefont {D.~L.}\ \bibnamefont
  {Mills}}\ and\ \bibinfo {author} {\bibfnamefont {V.~M.}\ \bibnamefont
  {Agranovich}},\ }\href@noop {} {\emph {\bibinfo {title} {Surface Polaritons:
  Electromagnetic Waves at Surfaces and Interfaces}}}\ (\bibinfo  {publisher}
  {North-Holland publ.},\ \bibinfo {year} {1982})\BibitemShut {NoStop}%
\bibitem [{\citenamefont {Mulvaney}(1996)}]{mulvaney1996Lengmuir}%
  \BibitemOpen
  \bibfield  {author} {\bibinfo {author} {\bibfnamefont {P.}~\bibnamefont
  {Mulvaney}},\ }\href@noop {} {\bibfield  {journal} {\bibinfo  {journal}
  {Langmuir}\ }\textbf {\bibinfo {volume} {12}},\ \bibinfo {pages} {788}
  (\bibinfo {year} {1996})}\BibitemShut {NoStop}%
\bibitem [{\citenamefont {Eberlein}\ \emph {et~al.}(2008)\citenamefont
  {Eberlein}, \citenamefont {Bangert}, \citenamefont {Nair}, \citenamefont
  {Jones}, \citenamefont {Gass}, \citenamefont {Bleloch}, \citenamefont
  {Novoselov}, \citenamefont {Geim},\ and\ \citenamefont
  {Briddon}}]{eberlein2008PRB}%
  \BibitemOpen
  \bibfield  {author} {\bibinfo {author} {\bibfnamefont {T.}~\bibnamefont
  {Eberlein}}, \bibinfo {author} {\bibfnamefont {U.}~\bibnamefont {Bangert}},
  \bibinfo {author} {\bibfnamefont {R.}~\bibnamefont {Nair}}, \bibinfo {author}
  {\bibfnamefont {R.}~\bibnamefont {Jones}}, \bibinfo {author} {\bibfnamefont
  {M.}~\bibnamefont {Gass}}, \bibinfo {author} {\bibfnamefont {A.}~\bibnamefont
  {Bleloch}}, \bibinfo {author} {\bibfnamefont {K.}~\bibnamefont {Novoselov}},
  \bibinfo {author} {\bibfnamefont {A.}~\bibnamefont {Geim}}, \ and\ \bibinfo
  {author} {\bibfnamefont {P.}~\bibnamefont {Briddon}},\ }\href@noop {}
  {\bibfield  {journal} {\bibinfo  {journal} {Phys. Rev. B}\ }\textbf {\bibinfo
  {volume} {77}},\ \bibinfo {pages} {233406} (\bibinfo {year}
  {2008})}\BibitemShut {NoStop}%
\bibitem [{\citenamefont {Tanaka}\ \emph {et~al.}(2013)\citenamefont {Tanaka},
  \citenamefont {Gomi}, \citenamefont {Funasaka}, \citenamefont {Asakawa},
  \citenamefont {Nakanishi},\ and\ \citenamefont
  {Moriwaki}}]{tanaka2013development}%
  \BibitemOpen
  \bibfield  {author} {\bibinfo {author} {\bibfnamefont {R.}~\bibnamefont
  {Tanaka}}, \bibinfo {author} {\bibfnamefont {R.}~\bibnamefont {Gomi}},
  \bibinfo {author} {\bibfnamefont {K.}~\bibnamefont {Funasaka}}, \bibinfo
  {author} {\bibfnamefont {D.}~\bibnamefont {Asakawa}}, \bibinfo {author}
  {\bibfnamefont {H.}~\bibnamefont {Nakanishi}}, \ and\ \bibinfo {author}
  {\bibfnamefont {H.}~\bibnamefont {Moriwaki}},\ }\href@noop {} {\bibfield
  {journal} {\bibinfo  {journal} {Analyst}\ }\textbf {\bibinfo {volume}
  {138}},\ \bibinfo {pages} {5437} (\bibinfo {year} {2013})}\BibitemShut
  {NoStop}%
\bibitem [{\citenamefont {Jeanmaire}\ and\ \citenamefont
  {Van~Duyne}(1977)}]{jeanmaire1977surface}%
  \BibitemOpen
  \bibfield  {author} {\bibinfo {author} {\bibfnamefont {D.~L.}\ \bibnamefont
  {Jeanmaire}}\ and\ \bibinfo {author} {\bibfnamefont {R.~P.}\ \bibnamefont
  {Van~Duyne}},\ }\href@noop {} {\bibfield  {journal} {\bibinfo  {journal} {J.
  Electroanalytical Chemistry and Interfacial Electrochemistry}\ }\textbf
  {\bibinfo {volume} {84}},\ \bibinfo {pages} {1} (\bibinfo {year}
  {1977})}\BibitemShut {NoStop}%
\bibitem [{\citenamefont {Otto}(1978)}]{otto1978SurfSci}%
  \BibitemOpen
  \bibfield  {author} {\bibinfo {author} {\bibfnamefont {A.}~\bibnamefont
  {Otto}},\ }\href@noop {} {\bibfield  {journal} {\bibinfo  {journal} {Surface
  Science}\ }\textbf {\bibinfo {volume} {75}},\ \bibinfo {pages} {L392}
  (\bibinfo {year} {1978})}\BibitemShut {NoStop}%
\bibitem [{\citenamefont {Eesley}(1981)}]{eesley1981PRB}%
  \BibitemOpen
  \bibfield  {author} {\bibinfo {author} {\bibfnamefont {G.}~\bibnamefont
  {Eesley}},\ }\href@noop {} {\bibfield  {journal} {\bibinfo  {journal} {Phys.
  Rev. B}\ }\textbf {\bibinfo {volume} {24}},\ \bibinfo {pages} {5477}
  (\bibinfo {year} {1981})}\BibitemShut {NoStop}%
\bibitem [{\citenamefont {Tsang}\ \emph {et~al.}(1979)\citenamefont {Tsang},
  \citenamefont {Kirtley},\ and\ \citenamefont {Bradley}}]{tsang1979PRL}%
  \BibitemOpen
  \bibfield  {author} {\bibinfo {author} {\bibfnamefont {J.}~\bibnamefont
  {Tsang}}, \bibinfo {author} {\bibfnamefont {J.}~\bibnamefont {Kirtley}}, \
  and\ \bibinfo {author} {\bibfnamefont {J.}~\bibnamefont {Bradley}},\
  }\href@noop {} {\bibfield  {journal} {\bibinfo  {journal} {Phys. Rev. Lett.}\
  }\textbf {\bibinfo {volume} {43}},\ \bibinfo {pages} {772} (\bibinfo {year}
  {1979})}\BibitemShut {NoStop}%
\bibitem [{\citenamefont {Fang}\ \emph {et~al.}(2013)\citenamefont {Fang},
  \citenamefont {Brodoceanu}, \citenamefont {Kraus},\ and\ \citenamefont
  {Voelcker}}]{fang2013RSC}%
  \BibitemOpen
  \bibfield  {author} {\bibinfo {author} {\bibfnamefont {C.}~\bibnamefont
  {Fang}}, \bibinfo {author} {\bibfnamefont {D.}~\bibnamefont {Brodoceanu}},
  \bibinfo {author} {\bibfnamefont {T.}~\bibnamefont {Kraus}}, \ and\ \bibinfo
  {author} {\bibfnamefont {N.~H.}\ \bibnamefont {Voelcker}},\ }\href@noop {}
  {\bibfield  {journal} {\bibinfo  {journal} {Rsc Advances}\ }\textbf {\bibinfo
  {volume} {3}},\ \bibinfo {pages} {4288} (\bibinfo {year} {2013})}\BibitemShut
  {NoStop}%
\bibitem [{\citenamefont {Cao}\ \emph {et~al.}(2002)\citenamefont {Cao},
  \citenamefont {Jin},\ and\ \citenamefont {Mirkin}}]{Cao2002science}%
  \BibitemOpen
  \bibfield  {author} {\bibinfo {author} {\bibfnamefont {Y.~C.}\ \bibnamefont
  {Cao}}, \bibinfo {author} {\bibfnamefont {R.}~\bibnamefont {Jin}}, \ and\
  \bibinfo {author} {\bibfnamefont {C.~A.}\ \bibnamefont {Mirkin}},\ }\href
  {\doibase 10.1126/science.297.5586.1536} {\bibfield  {journal} {\bibinfo
  {journal} {Science}\ }\textbf {\bibinfo {volume} {297}},\ \bibinfo {pages}
  {1536} (\bibinfo {year} {2002})}\BibitemShut {NoStop}%
\bibitem [{\citenamefont {Ang}\ \emph {et~al.}(2015)\citenamefont {Ang},
  \citenamefont {Thevarajah}, \citenamefont {Alias},\ and\ \citenamefont
  {Khor}}]{Ang2015hemoglobin}%
  \BibitemOpen
  \bibfield  {author} {\bibinfo {author} {\bibfnamefont {S.~H.}\ \bibnamefont
  {Ang}}, \bibinfo {author} {\bibfnamefont {M.}~\bibnamefont {Thevarajah}},
  \bibinfo {author} {\bibfnamefont {Y.}~\bibnamefont {Alias}}, \ and\ \bibinfo
  {author} {\bibfnamefont {S.~M.}\ \bibnamefont {Khor}},\ }\href {\doibase
  http://dx.doi.org/10.1016/j.cca.2014.10.019} {\bibfield  {journal} {\bibinfo
  {journal} {Clinica Chimica Acta}\ }\textbf {\bibinfo {volume} {439}},\
  \bibinfo {pages} {202 } (\bibinfo {year} {2015})}\BibitemShut {NoStop}%
\bibitem [{\citenamefont {Hawrylak}\ and\ \citenamefont
  {Quinn}(1986)}]{hawrylak1986APL}%
  \BibitemOpen
  \bibfield  {author} {\bibinfo {author} {\bibfnamefont {P.}~\bibnamefont
  {Hawrylak}}\ and\ \bibinfo {author} {\bibfnamefont {J.~J.}\ \bibnamefont
  {Quinn}},\ }\href@noop {} {\bibfield  {journal} {\bibinfo  {journal} {Appl.
  Phys. Lett.}\ }\textbf {\bibinfo {volume} {49}},\ \bibinfo {pages} {280}
  (\bibinfo {year} {1986})}\BibitemShut {NoStop}%
\bibitem [{\citenamefont {Kempa}\ \emph {et~al.}(1988)\citenamefont {Kempa},
  \citenamefont {Bakshi},\ and\ \citenamefont {Cen}}]{kempa1988}%
  \BibitemOpen
  \bibfield  {author} {\bibinfo {author} {\bibfnamefont {K.}~\bibnamefont
  {Kempa}}, \bibinfo {author} {\bibfnamefont {P.}~\bibnamefont {Bakshi}}, \
  and\ \bibinfo {author} {\bibfnamefont {J.}~\bibnamefont {Cen}},\ }in\
  \href@noop {} {\emph {\bibinfo {booktitle} {1988 Semiconductor Symposium}}}\
  (\bibinfo {organization} {International Society for Optics and Photonics},\
  \bibinfo {year} {1988})\ pp.\ \bibinfo {pages} {62--67}\BibitemShut {NoStop}%
\bibitem [{\citenamefont {Maier}(2006)}]{maier2006optcomm}%
  \BibitemOpen
  \bibfield  {author} {\bibinfo {author} {\bibfnamefont {S.~A.}\ \bibnamefont
  {Maier}},\ }\href@noop {} {\bibfield  {journal} {\bibinfo  {journal} {Optics
  communications}\ }\textbf {\bibinfo {volume} {258}},\ \bibinfo {pages} {295}
  (\bibinfo {year} {2006})}\BibitemShut {NoStop}%
\bibitem [{\citenamefont {Sirtori}\ \emph {et~al.}(1998)\citenamefont
  {Sirtori}, \citenamefont {Gmachl}, \citenamefont {Capasso}, \citenamefont
  {Faist}, \citenamefont {Sivco}, \citenamefont {Hutchinson},\ and\
  \citenamefont {Cho}}]{sirtori1998OptLett}%
  \BibitemOpen
  \bibfield  {author} {\bibinfo {author} {\bibfnamefont {C.}~\bibnamefont
  {Sirtori}}, \bibinfo {author} {\bibfnamefont {C.}~\bibnamefont {Gmachl}},
  \bibinfo {author} {\bibfnamefont {F.}~\bibnamefont {Capasso}}, \bibinfo
  {author} {\bibfnamefont {J.}~\bibnamefont {Faist}}, \bibinfo {author}
  {\bibfnamefont {D.~L.}\ \bibnamefont {Sivco}}, \bibinfo {author}
  {\bibfnamefont {A.~L.}\ \bibnamefont {Hutchinson}}, \ and\ \bibinfo {author}
  {\bibfnamefont {A.~Y.}\ \bibnamefont {Cho}},\ }\href@noop {} {\bibfield
  {journal} {\bibinfo  {journal} {Opt. Lett.}\ }\textbf {\bibinfo {volume}
  {23}},\ \bibinfo {pages} {1366} (\bibinfo {year} {1998})}\BibitemShut
  {NoStop}%
\bibitem [{\citenamefont {Tredicucci}\ \emph {et~al.}(2000)\citenamefont
  {Tredicucci}, \citenamefont {Gmachl}, \citenamefont {Capasso}, \citenamefont
  {Hutchinson}, \citenamefont {Sivco},\ and\ \citenamefont
  {Cho}}]{tredicucci2000APL}%
  \BibitemOpen
  \bibfield  {author} {\bibinfo {author} {\bibfnamefont {A.}~\bibnamefont
  {Tredicucci}}, \bibinfo {author} {\bibfnamefont {C.}~\bibnamefont {Gmachl}},
  \bibinfo {author} {\bibfnamefont {F.}~\bibnamefont {Capasso}}, \bibinfo
  {author} {\bibfnamefont {A.~F.}\ \bibnamefont {Hutchinson}}, \bibinfo
  {author} {\bibfnamefont {D.~L.}\ \bibnamefont {Sivco}}, \ and\ \bibinfo
  {author} {\bibfnamefont {A.~Y.}\ \bibnamefont {Cho}},\ }\href@noop {}
  {\bibfield  {journal} {\bibinfo  {journal} {Appl. Phys. Lett.}\ }\textbf
  {\bibinfo {volume} {76}},\ \bibinfo {pages} {2164} (\bibinfo {year}
  {2000})}\BibitemShut {NoStop}%
\bibitem [{\citenamefont {Babuty}\ \emph {et~al.}(2010)\citenamefont {Babuty},
  \citenamefont {Bousseksou}, \citenamefont {Tetienne}, \citenamefont {Doyen},
  \citenamefont {Sirtori}, \citenamefont {Beaudoin}, \citenamefont {Sagnes},
  \citenamefont {De~Wilde},\ and\ \citenamefont {Colombelli}}]{babuty2010PRL}%
  \BibitemOpen
  \bibfield  {author} {\bibinfo {author} {\bibfnamefont {A.}~\bibnamefont
  {Babuty}}, \bibinfo {author} {\bibfnamefont {A.}~\bibnamefont {Bousseksou}},
  \bibinfo {author} {\bibfnamefont {J.-P.}\ \bibnamefont {Tetienne}}, \bibinfo
  {author} {\bibfnamefont {I.~M.}\ \bibnamefont {Doyen}}, \bibinfo {author}
  {\bibfnamefont {C.}~\bibnamefont {Sirtori}}, \bibinfo {author} {\bibfnamefont
  {G.}~\bibnamefont {Beaudoin}}, \bibinfo {author} {\bibfnamefont
  {I.}~\bibnamefont {Sagnes}}, \bibinfo {author} {\bibfnamefont
  {Y.}~\bibnamefont {De~Wilde}}, \ and\ \bibinfo {author} {\bibfnamefont
  {R.}~\bibnamefont {Colombelli}},\ }\href@noop {} {\bibfield  {journal}
  {\bibinfo  {journal} {Phys. Rev. Lett.}\ }\textbf {\bibinfo {volume} {104}},\
  \bibinfo {pages} {226806} (\bibinfo {year} {2010})}\BibitemShut {NoStop}%
\bibitem [{\citenamefont {Oulton}\ \emph {et~al.}(2009)\citenamefont {Oulton},
  \citenamefont {Sorger}, \citenamefont {Zentgraf}, \citenamefont {Ma},
  \citenamefont {Gladden}, \citenamefont {Dai}, \citenamefont {Bartal},\ and\
  \citenamefont {Zhang}}]{oulton2009nature}%
  \BibitemOpen
  \bibfield  {author} {\bibinfo {author} {\bibfnamefont {R.~F.}\ \bibnamefont
  {Oulton}}, \bibinfo {author} {\bibfnamefont {V.~J.}\ \bibnamefont {Sorger}},
  \bibinfo {author} {\bibfnamefont {T.}~\bibnamefont {Zentgraf}}, \bibinfo
  {author} {\bibfnamefont {R.-M.}\ \bibnamefont {Ma}}, \bibinfo {author}
  {\bibfnamefont {C.}~\bibnamefont {Gladden}}, \bibinfo {author} {\bibfnamefont
  {L.}~\bibnamefont {Dai}}, \bibinfo {author} {\bibfnamefont {G.}~\bibnamefont
  {Bartal}}, \ and\ \bibinfo {author} {\bibfnamefont {X.}~\bibnamefont
  {Zhang}},\ }\href@noop {} {\bibfield  {journal} {\bibinfo  {journal}
  {Nature}\ }\textbf {\bibinfo {volume} {461}},\ \bibinfo {pages} {629}
  (\bibinfo {year} {2009})}\BibitemShut {NoStop}%
\bibitem [{\citenamefont {Noginov}\ \emph {et~al.}(2008)\citenamefont
  {Noginov}, \citenamefont {Zhu}, \citenamefont {Mayy}, \citenamefont {Ritzo},
  \citenamefont {Noginova},\ and\ \citenamefont {Podolskiy}}]{noginov2008PRL}%
  \BibitemOpen
  \bibfield  {author} {\bibinfo {author} {\bibfnamefont {M.~A.}\ \bibnamefont
  {Noginov}}, \bibinfo {author} {\bibfnamefont {G.}~\bibnamefont {Zhu}},
  \bibinfo {author} {\bibfnamefont {M.}~\bibnamefont {Mayy}}, \bibinfo {author}
  {\bibfnamefont {B.~A.}\ \bibnamefont {Ritzo}}, \bibinfo {author}
  {\bibfnamefont {N.}~\bibnamefont {Noginova}}, \ and\ \bibinfo {author}
  {\bibfnamefont {V.~A.}\ \bibnamefont {Podolskiy}},\ }\href {\doibase
  10.1103/PhysRevLett.101.226806} {\bibfield  {journal} {\bibinfo  {journal}
  {Phys. Rev. Lett.}\ }\textbf {\bibinfo {volume} {101}},\ \bibinfo {pages}
  {226806} (\bibinfo {year} {2008})}\BibitemShut {NoStop}%
\bibitem [{\citenamefont {Bergman}\ and\ \citenamefont
  {Stockman}(2003)}]{bergman2003PRL}%
  \BibitemOpen
  \bibfield  {author} {\bibinfo {author} {\bibfnamefont {D.~J.}\ \bibnamefont
  {Bergman}}\ and\ \bibinfo {author} {\bibfnamefont {M.~I.}\ \bibnamefont
  {Stockman}},\ }\href@noop {} {\bibfield  {journal} {\bibinfo  {journal}
  {Phys. Rev. Lett.}\ }\textbf {\bibinfo {volume} {90}},\ \bibinfo {pages}
  {027402} (\bibinfo {year} {2003})}\BibitemShut {NoStop}%
\bibitem [{\citenamefont {Protsenko}\ \emph {et~al.}(2005)\citenamefont
  {Protsenko}, \citenamefont {Uskov}, \citenamefont {Zaimidoroga},
  \citenamefont {Samoilov},\ and\ \citenamefont {O'reilly}}]{protsenko2005PRA}%
  \BibitemOpen
  \bibfield  {author} {\bibinfo {author} {\bibfnamefont {I.~E.}\ \bibnamefont
  {Protsenko}}, \bibinfo {author} {\bibfnamefont {A.~V.}\ \bibnamefont
  {Uskov}}, \bibinfo {author} {\bibfnamefont {O.}~\bibnamefont {Zaimidoroga}},
  \bibinfo {author} {\bibfnamefont {V.}~\bibnamefont {Samoilov}}, \ and\
  \bibinfo {author} {\bibfnamefont {E.}~\bibnamefont {O'reilly}},\ }\href@noop
  {} {\bibfield  {journal} {\bibinfo  {journal} {Phys. Rev. A}\ }\textbf
  {\bibinfo {volume} {71}},\ \bibinfo {pages} {063812} (\bibinfo {year}
  {2005})}\BibitemShut {NoStop}%
\bibitem [{\citenamefont {Andrianov}\ \emph {et~al.}(2012)\citenamefont
  {Andrianov}, \citenamefont {Pukhov}, \citenamefont {Dorofeenko},
  \citenamefont {Vinogradov},\ and\ \citenamefont
  {Lisyansky}}]{andrianov2012PRB}%
  \BibitemOpen
  \bibfield  {author} {\bibinfo {author} {\bibfnamefont {E.}~\bibnamefont
  {Andrianov}}, \bibinfo {author} {\bibfnamefont {A.}~\bibnamefont {Pukhov}},
  \bibinfo {author} {\bibfnamefont {A.}~\bibnamefont {Dorofeenko}}, \bibinfo
  {author} {\bibfnamefont {A.}~\bibnamefont {Vinogradov}}, \ and\ \bibinfo
  {author} {\bibfnamefont {A.}~\bibnamefont {Lisyansky}},\ }\href@noop {}
  {\bibfield  {journal} {\bibinfo  {journal} {Phys. Rev. B}\ }\textbf {\bibinfo
  {volume} {85}},\ \bibinfo {pages} {165419} (\bibinfo {year}
  {2012})}\BibitemShut {NoStop}%
\bibitem [{\citenamefont {Andrianov}\ \emph
  {et~al.}(2011{\natexlab{a}})\citenamefont {Andrianov}, \citenamefont
  {Pukhov}, \citenamefont {Dorofeenko}, \citenamefont {Vinogradov},\ and\
  \citenamefont {Lisyansky}}]{andrianov2011OptExp}%
  \BibitemOpen
  \bibfield  {author} {\bibinfo {author} {\bibfnamefont {E.}~\bibnamefont
  {Andrianov}}, \bibinfo {author} {\bibfnamefont {A.}~\bibnamefont {Pukhov}},
  \bibinfo {author} {\bibfnamefont {A.}~\bibnamefont {Dorofeenko}}, \bibinfo
  {author} {\bibfnamefont {A.}~\bibnamefont {Vinogradov}}, \ and\ \bibinfo
  {author} {\bibfnamefont {A.}~\bibnamefont {Lisyansky}},\ }\href@noop {}
  {\bibfield  {journal} {\bibinfo  {journal} {Opt. Express}\ }\textbf {\bibinfo
  {volume} {19}},\ \bibinfo {pages} {24849} (\bibinfo {year}
  {2011}{\natexlab{a}})}\BibitemShut {NoStop}%
\bibitem [{\citenamefont {Andrianov}\ \emph
  {et~al.}(2011{\natexlab{b}})\citenamefont {Andrianov}, \citenamefont
  {Pukhov}, \citenamefont {Dorofeenko}, \citenamefont {Vinogradov},\ and\
  \citenamefont {Lisyansky}}]{andrianov2011OptLett}%
  \BibitemOpen
  \bibfield  {author} {\bibinfo {author} {\bibfnamefont {E.}~\bibnamefont
  {Andrianov}}, \bibinfo {author} {\bibfnamefont {A.}~\bibnamefont {Pukhov}},
  \bibinfo {author} {\bibfnamefont {A.}~\bibnamefont {Dorofeenko}}, \bibinfo
  {author} {\bibfnamefont {A.}~\bibnamefont {Vinogradov}}, \ and\ \bibinfo
  {author} {\bibfnamefont {A.}~\bibnamefont {Lisyansky}},\ }\href@noop {}
  {\bibfield  {journal} {\bibinfo  {journal} {Opt. Lett.}\ }\textbf {\bibinfo
  {volume} {36}},\ \bibinfo {pages} {4302} (\bibinfo {year}
  {2011}{\natexlab{b}})}\BibitemShut {NoStop}%
\bibitem [{\citenamefont {Noginov}\ \emph {et~al.}(2009)\citenamefont
  {Noginov}, \citenamefont {Zhu}, \citenamefont {Belgrave}, \citenamefont
  {Bakker}, \citenamefont {Shalaev}, \citenamefont {Narimanov}, \citenamefont
  {Stout}, \citenamefont {Herz}, \citenamefont {Suteewong},\ and\ \citenamefont
  {Wiesner}}]{noginov2009Nature}%
  \BibitemOpen
  \bibfield  {author} {\bibinfo {author} {\bibfnamefont {M.}~\bibnamefont
  {Noginov}}, \bibinfo {author} {\bibfnamefont {G.}~\bibnamefont {Zhu}},
  \bibinfo {author} {\bibfnamefont {A.}~\bibnamefont {Belgrave}}, \bibinfo
  {author} {\bibfnamefont {R.}~\bibnamefont {Bakker}}, \bibinfo {author}
  {\bibfnamefont {V.}~\bibnamefont {Shalaev}}, \bibinfo {author} {\bibfnamefont
  {E.}~\bibnamefont {Narimanov}}, \bibinfo {author} {\bibfnamefont
  {S.}~\bibnamefont {Stout}}, \bibinfo {author} {\bibfnamefont
  {E.}~\bibnamefont {Herz}}, \bibinfo {author} {\bibfnamefont {T.}~\bibnamefont
  {Suteewong}}, \ and\ \bibinfo {author} {\bibfnamefont {U.}~\bibnamefont
  {Wiesner}},\ }\href@noop {} {\bibfield  {journal} {\bibinfo  {journal}
  {Nature}\ }\textbf {\bibinfo {volume} {460}},\ \bibinfo {pages} {1110}
  (\bibinfo {year} {2009})}\BibitemShut {NoStop}%
\bibitem [{\citenamefont {Lu}\ \emph {et~al.}(2012)\citenamefont {Lu},
  \citenamefont {Kim}, \citenamefont {Chen}, \citenamefont {Wu}, \citenamefont
  {Dabidian}, \citenamefont {Sanders}, \citenamefont {Wang}, \citenamefont
  {Lu}, \citenamefont {Li}, \citenamefont {Qiu} \emph
  {et~al.}}]{lu2012Science}%
  \BibitemOpen
  \bibfield  {author} {\bibinfo {author} {\bibfnamefont {Y.-J.}\ \bibnamefont
  {Lu}}, \bibinfo {author} {\bibfnamefont {J.}~\bibnamefont {Kim}}, \bibinfo
  {author} {\bibfnamefont {H.-Y.}\ \bibnamefont {Chen}}, \bibinfo {author}
  {\bibfnamefont {C.}~\bibnamefont {Wu}}, \bibinfo {author} {\bibfnamefont
  {N.}~\bibnamefont {Dabidian}}, \bibinfo {author} {\bibfnamefont {C.~E.}\
  \bibnamefont {Sanders}}, \bibinfo {author} {\bibfnamefont {C.-Y.}\
  \bibnamefont {Wang}}, \bibinfo {author} {\bibfnamefont {M.-Y.}\ \bibnamefont
  {Lu}}, \bibinfo {author} {\bibfnamefont {B.-H.}\ \bibnamefont {Li}}, \bibinfo
  {author} {\bibfnamefont {X.}~\bibnamefont {Qiu}},  \emph {et~al.},\
  }\href@noop {} {\bibfield  {journal} {\bibinfo  {journal} {Science}\ }\textbf
  {\bibinfo {volume} {337}},\ \bibinfo {pages} {450} (\bibinfo {year}
  {2012})}\BibitemShut {NoStop}%
\bibitem [{\citenamefont {Hill}\ \emph {et~al.}(2009)\citenamefont {Hill},
  \citenamefont {Marell}, \citenamefont {Leong}, \citenamefont {Smalbrugge},
  \citenamefont {Zhu}, \citenamefont {Sun}, \citenamefont {van Veldhoven},
  \citenamefont {Geluk}, \citenamefont {Karouta}, \citenamefont {Oei} \emph
  {et~al.}}]{hill2009OptExp}%
  \BibitemOpen
  \bibfield  {author} {\bibinfo {author} {\bibfnamefont {M.~T.}\ \bibnamefont
  {Hill}}, \bibinfo {author} {\bibfnamefont {M.}~\bibnamefont {Marell}},
  \bibinfo {author} {\bibfnamefont {E.~S.}\ \bibnamefont {Leong}}, \bibinfo
  {author} {\bibfnamefont {B.}~\bibnamefont {Smalbrugge}}, \bibinfo {author}
  {\bibfnamefont {Y.}~\bibnamefont {Zhu}}, \bibinfo {author} {\bibfnamefont
  {M.}~\bibnamefont {Sun}}, \bibinfo {author} {\bibfnamefont {P.~J.}\
  \bibnamefont {van Veldhoven}}, \bibinfo {author} {\bibfnamefont {E.~J.}\
  \bibnamefont {Geluk}}, \bibinfo {author} {\bibfnamefont {F.}~\bibnamefont
  {Karouta}}, \bibinfo {author} {\bibfnamefont {Y.-S.}\ \bibnamefont {Oei}},
  \emph {et~al.},\ }\href@noop {} {\bibfield  {journal} {\bibinfo  {journal}
  {Opt. Express}\ }\textbf {\bibinfo {volume} {17}},\ \bibinfo {pages} {11107}
  (\bibinfo {year} {2009})}\BibitemShut {NoStop}%
\bibitem [{\citenamefont {Suh}\ \emph {et~al.}(2012)\citenamefont {Suh},
  \citenamefont {Kim}, \citenamefont {Zhou}, \citenamefont {Huntington},
  \citenamefont {Co}, \citenamefont {Wasielewski},\ and\ \citenamefont
  {Odom}}]{suh2012NanoLett}%
  \BibitemOpen
  \bibfield  {author} {\bibinfo {author} {\bibfnamefont {J.~Y.}\ \bibnamefont
  {Suh}}, \bibinfo {author} {\bibfnamefont {C.~H.}\ \bibnamefont {Kim}},
  \bibinfo {author} {\bibfnamefont {W.}~\bibnamefont {Zhou}}, \bibinfo {author}
  {\bibfnamefont {M.~D.}\ \bibnamefont {Huntington}}, \bibinfo {author}
  {\bibfnamefont {D.~T.}\ \bibnamefont {Co}}, \bibinfo {author} {\bibfnamefont
  {M.~R.}\ \bibnamefont {Wasielewski}}, \ and\ \bibinfo {author} {\bibfnamefont
  {T.~W.}\ \bibnamefont {Odom}},\ }\href@noop {} {\bibfield  {journal}
  {\bibinfo  {journal} {Nano Lett.}\ }\textbf {\bibinfo {volume} {12}},\
  \bibinfo {pages} {5769} (\bibinfo {year} {2012})}\BibitemShut {NoStop}%
\bibitem [{\citenamefont {van Beijnum}\ \emph {et~al.}(2013)\citenamefont {van
  Beijnum}, \citenamefont {van Veldhoven}, \citenamefont {Geluk}, \citenamefont
  {de~Dood}, \citenamefont {Gert},\ and\ \citenamefont {van
  Exter}}]{van2013PRL}%
  \BibitemOpen
  \bibfield  {author} {\bibinfo {author} {\bibfnamefont {F.}~\bibnamefont {van
  Beijnum}}, \bibinfo {author} {\bibfnamefont {P.~J.}\ \bibnamefont {van
  Veldhoven}}, \bibinfo {author} {\bibfnamefont {E.~J.}\ \bibnamefont {Geluk}},
  \bibinfo {author} {\bibfnamefont {M.~J.}\ \bibnamefont {de~Dood}}, \bibinfo
  {author} {\bibfnamefont {W.}~\bibnamefont {Gert}}, \ and\ \bibinfo {author}
  {\bibfnamefont {M.~P.}\ \bibnamefont {van Exter}},\ }\href@noop {} {\bibfield
   {journal} {\bibinfo  {journal} {Phys. Rev. Lett.}\ }\textbf {\bibinfo
  {volume} {110}},\ \bibinfo {pages} {206802} (\bibinfo {year}
  {2013})}\BibitemShut {NoStop}%
\bibitem [{\citenamefont {Berini}\ and\ \citenamefont
  {De~Leon}(2012)}]{berini2012NatPhot}%
  \BibitemOpen
  \bibfield  {author} {\bibinfo {author} {\bibfnamefont {P.}~\bibnamefont
  {Berini}}\ and\ \bibinfo {author} {\bibfnamefont {I.}~\bibnamefont
  {De~Leon}},\ }\href@noop {} {\bibfield  {journal} {\bibinfo  {journal} {Nat.
  Photonics}\ }\textbf {\bibinfo {volume} {6}},\ \bibinfo {pages} {16}
  (\bibinfo {year} {2012})}\BibitemShut {NoStop}%
\bibitem [{\citenamefont {Tame}\ \emph {et~al.}(2013)\citenamefont {Tame},
  \citenamefont {McEnery}, \citenamefont {{\"O}zdemir}, \citenamefont {Lee},
  \citenamefont {Maier},\ and\ \citenamefont {Kim}}]{tame2013NatPhys}%
  \BibitemOpen
  \bibfield  {author} {\bibinfo {author} {\bibfnamefont {M.}~\bibnamefont
  {Tame}}, \bibinfo {author} {\bibfnamefont {K.}~\bibnamefont {McEnery}},
  \bibinfo {author} {\bibfnamefont {{\c{S}}.}~\bibnamefont {{\"O}zdemir}},
  \bibinfo {author} {\bibfnamefont {J.}~\bibnamefont {Lee}}, \bibinfo {author}
  {\bibfnamefont {S.}~\bibnamefont {Maier}}, \ and\ \bibinfo {author}
  {\bibfnamefont {M.}~\bibnamefont {Kim}},\ }\href@noop {} {\bibfield
  {journal} {\bibinfo  {journal} {Nat. Physics}\ }\textbf {\bibinfo {volume}
  {9}},\ \bibinfo {pages} {329} (\bibinfo {year} {2013})}\BibitemShut {NoStop}%
\bibitem [{\citenamefont {Ju}\ \emph {et~al.}(2011{\natexlab{a}})\citenamefont
  {Ju}, \citenamefont {Geng}, \citenamefont {Horng}, \citenamefont {Girit},
  \citenamefont {Martin}, \citenamefont {Hao}, \citenamefont {Bechtel},
  \citenamefont {Liang}, \citenamefont {Zettl}, \citenamefont {Shen} \emph
  {et~al.}}]{jacob2011Science}%
  \BibitemOpen
  \bibfield  {author} {\bibinfo {author} {\bibfnamefont {L.}~\bibnamefont
  {Ju}}, \bibinfo {author} {\bibfnamefont {B.}~\bibnamefont {Geng}}, \bibinfo
  {author} {\bibfnamefont {J.}~\bibnamefont {Horng}}, \bibinfo {author}
  {\bibfnamefont {C.}~\bibnamefont {Girit}}, \bibinfo {author} {\bibfnamefont
  {M.}~\bibnamefont {Martin}}, \bibinfo {author} {\bibfnamefont
  {Z.}~\bibnamefont {Hao}}, \bibinfo {author} {\bibfnamefont {H.~A.}\
  \bibnamefont {Bechtel}}, \bibinfo {author} {\bibfnamefont {X.}~\bibnamefont
  {Liang}}, \bibinfo {author} {\bibfnamefont {A.}~\bibnamefont {Zettl}},
  \bibinfo {author} {\bibfnamefont {Y.~R.}\ \bibnamefont {Shen}},  \emph
  {et~al.},\ }\href@noop {} {\bibfield  {journal} {\bibinfo  {journal}
  {Science}\ }\textbf {\bibinfo {volume} {334}},\ \bibinfo {pages} {463}
  (\bibinfo {year} {2011}{\natexlab{a}})}\BibitemShut {NoStop}%
\bibitem [{\citenamefont {Ju}\ \emph {et~al.}(2011{\natexlab{b}})\citenamefont
  {Ju}, \citenamefont {Geng}, \citenamefont {Horng}, \citenamefont {Girit},
  \citenamefont {Martin}, \citenamefont {Hao}, \citenamefont {Bechtel},
  \citenamefont {Liang}, \citenamefont {Zettl}, \citenamefont {Shen} \emph
  {et~al.}}]{ju2011NatNano}%
  \BibitemOpen
  \bibfield  {author} {\bibinfo {author} {\bibfnamefont {L.}~\bibnamefont
  {Ju}}, \bibinfo {author} {\bibfnamefont {B.}~\bibnamefont {Geng}}, \bibinfo
  {author} {\bibfnamefont {J.}~\bibnamefont {Horng}}, \bibinfo {author}
  {\bibfnamefont {C.}~\bibnamefont {Girit}}, \bibinfo {author} {\bibfnamefont
  {M.}~\bibnamefont {Martin}}, \bibinfo {author} {\bibfnamefont
  {Z.}~\bibnamefont {Hao}}, \bibinfo {author} {\bibfnamefont {H.~A.}\
  \bibnamefont {Bechtel}}, \bibinfo {author} {\bibfnamefont {X.}~\bibnamefont
  {Liang}}, \bibinfo {author} {\bibfnamefont {A.}~\bibnamefont {Zettl}},
  \bibinfo {author} {\bibfnamefont {Y.~R.}\ \bibnamefont {Shen}},  \emph
  {et~al.},\ }\href@noop {} {\bibfield  {journal} {\bibinfo  {journal} {Nat.
  Nanotechnology}\ }\textbf {\bibinfo {volume} {6}},\ \bibinfo {pages} {630}
  (\bibinfo {year} {2011}{\natexlab{b}})}\BibitemShut {NoStop}%
\bibitem [{\citenamefont {Koppens}\ \emph {et~al.}(2011)\citenamefont
  {Koppens}, \citenamefont {Chang},\ and\ \citenamefont {Garcia~de
  Abajo}}]{koppens2011NanoLett}%
  \BibitemOpen
  \bibfield  {author} {\bibinfo {author} {\bibfnamefont {F.~H.}\ \bibnamefont
  {Koppens}}, \bibinfo {author} {\bibfnamefont {D.~E.}\ \bibnamefont {Chang}},
  \ and\ \bibinfo {author} {\bibfnamefont {F.~J.}\ \bibnamefont {Garcia~de
  Abajo}},\ }\href@noop {} {\bibfield  {journal} {\bibinfo  {journal} {Nano
  Lett.}\ }\textbf {\bibinfo {volume} {11}},\ \bibinfo {pages} {3370} (\bibinfo
  {year} {2011})}\BibitemShut {NoStop}%
\bibitem [{\citenamefont {Grigorenko}\ \emph {et~al.}(2012)\citenamefont
  {Grigorenko}, \citenamefont {Polini},\ and\ \citenamefont
  {Novoselov}}]{grigorenko2012NatPhot}%
  \BibitemOpen
  \bibfield  {author} {\bibinfo {author} {\bibfnamefont {A.}~\bibnamefont
  {Grigorenko}}, \bibinfo {author} {\bibfnamefont {M.}~\bibnamefont {Polini}},
  \ and\ \bibinfo {author} {\bibfnamefont {K.}~\bibnamefont {Novoselov}},\
  }\href@noop {} {\bibfield  {journal} {\bibinfo  {journal} {Nat. Photonics}\
  }\textbf {\bibinfo {volume} {6}},\ \bibinfo {pages} {749} (\bibinfo {year}
  {2012})}\BibitemShut {NoStop}%
\bibitem [{\citenamefont {Garcia~de Abajo}(2014)}]{garcia2014ACSPhoton}%
  \BibitemOpen
  \bibfield  {author} {\bibinfo {author} {\bibfnamefont {F.~J.}\ \bibnamefont
  {Garcia~de Abajo}},\ }\href@noop {} {\bibfield  {journal} {\bibinfo
  {journal} {ACS Photonics}\ }\textbf {\bibinfo {volume} {1}},\ \bibinfo
  {pages} {135} (\bibinfo {year} {2014})}\BibitemShut {NoStop}%
\bibitem [{\citenamefont {Low}\ and\ \citenamefont
  {Avouris}(2014)}]{low2014ACSNano}%
  \BibitemOpen
  \bibfield  {author} {\bibinfo {author} {\bibfnamefont {T.}~\bibnamefont
  {Low}}\ and\ \bibinfo {author} {\bibfnamefont {P.}~\bibnamefont {Avouris}},\
  }\href@noop {} {\bibfield  {journal} {\bibinfo  {journal} {ACS Nano}\
  }\textbf {\bibinfo {volume} {8}},\ \bibinfo {pages} {1086} (\bibinfo {year}
  {2014})}\BibitemShut {NoStop}%
\bibitem [{\citenamefont {Maier}(2012)}]{maier2012NatPhys}%
  \BibitemOpen
  \bibfield  {author} {\bibinfo {author} {\bibfnamefont {S.~A.}\ \bibnamefont
  {Maier}},\ }\href@noop {} {\bibfield  {journal} {\bibinfo  {journal} {Nat.
  Phys.}\ }\textbf {\bibinfo {volume} {8}},\ \bibinfo {pages} {581} (\bibinfo
  {year} {2012})}\BibitemShut {NoStop}%
\bibitem [{\citenamefont {Novoselov}\ \emph {et~al.}(2005)\citenamefont
  {Novoselov}, \citenamefont {Geim}, \citenamefont {Morozov}, \citenamefont
  {Jiang}, \citenamefont {Katsnelson}, \citenamefont {Grigorieva},
  \citenamefont {Dubonos},\ and\ \citenamefont {Firsov}}]{novoselov2005Nature}%
  \BibitemOpen
  \bibfield  {author} {\bibinfo {author} {\bibfnamefont {K.}~\bibnamefont
  {Novoselov}}, \bibinfo {author} {\bibfnamefont {A.~K.}\ \bibnamefont {Geim}},
  \bibinfo {author} {\bibfnamefont {S.}~\bibnamefont {Morozov}}, \bibinfo
  {author} {\bibfnamefont {D.}~\bibnamefont {Jiang}}, \bibinfo {author}
  {\bibfnamefont {M.}~\bibnamefont {Katsnelson}}, \bibinfo {author}
  {\bibfnamefont {I.}~\bibnamefont {Grigorieva}}, \bibinfo {author}
  {\bibfnamefont {S.}~\bibnamefont {Dubonos}}, \ and\ \bibinfo {author}
  {\bibfnamefont {A.}~\bibnamefont {Firsov}},\ }\href@noop {} {\bibfield
  {journal} {\bibinfo  {journal} {Nature}\ }\textbf {\bibinfo {volume} {438}},\
  \bibinfo {pages} {197} (\bibinfo {year} {2005})}\BibitemShut {NoStop}%
\bibitem [{\citenamefont {Zhang}\ \emph {et~al.}(2005)\citenamefont {Zhang},
  \citenamefont {Tan}, \citenamefont {Stormer},\ and\ \citenamefont
  {Kim}}]{zhang2005Nature}%
  \BibitemOpen
  \bibfield  {author} {\bibinfo {author} {\bibfnamefont {Y.}~\bibnamefont
  {Zhang}}, \bibinfo {author} {\bibfnamefont {Y.-W.}\ \bibnamefont {Tan}},
  \bibinfo {author} {\bibfnamefont {H.~L.}\ \bibnamefont {Stormer}}, \ and\
  \bibinfo {author} {\bibfnamefont {P.}~\bibnamefont {Kim}},\ }\href@noop {}
  {\bibfield  {journal} {\bibinfo  {journal} {Nature}\ }\textbf {\bibinfo
  {volume} {438}},\ \bibinfo {pages} {201} (\bibinfo {year}
  {2005})}\BibitemShut {NoStop}%
\bibitem [{\citenamefont {Bolotin}\ \emph {et~al.}(2008)\citenamefont
  {Bolotin}, \citenamefont {Sikes}, \citenamefont {Jiang}, \citenamefont
  {Klima}, \citenamefont {Fudenberg}, \citenamefont {Hone}, \citenamefont
  {Kim},\ and\ \citenamefont {Stormer}}]{bolotin2008SolStCom}%
  \BibitemOpen
  \bibfield  {author} {\bibinfo {author} {\bibfnamefont {K.~I.}\ \bibnamefont
  {Bolotin}}, \bibinfo {author} {\bibfnamefont {K.}~\bibnamefont {Sikes}},
  \bibinfo {author} {\bibfnamefont {Z.}~\bibnamefont {Jiang}}, \bibinfo
  {author} {\bibfnamefont {M.}~\bibnamefont {Klima}}, \bibinfo {author}
  {\bibfnamefont {G.}~\bibnamefont {Fudenberg}}, \bibinfo {author}
  {\bibfnamefont {J.}~\bibnamefont {Hone}}, \bibinfo {author} {\bibfnamefont
  {P.}~\bibnamefont {Kim}}, \ and\ \bibinfo {author} {\bibfnamefont
  {H.}~\bibnamefont {Stormer}},\ }\href@noop {} {\bibfield  {journal} {\bibinfo
   {journal} {Solid State Comm.}\ }\textbf {\bibinfo {volume} {146}},\ \bibinfo
  {pages} {351} (\bibinfo {year} {2008})}\BibitemShut {NoStop}%
\bibitem [{\citenamefont {Balandin}\ \emph {et~al.}(2008)\citenamefont
  {Balandin}, \citenamefont {Ghosh}, \citenamefont {Bao}, \citenamefont
  {Calizo}, \citenamefont {Teweldebrhan}, \citenamefont {Miao},\ and\
  \citenamefont {Lau}}]{balandin2008NanoLett}%
  \BibitemOpen
  \bibfield  {author} {\bibinfo {author} {\bibfnamefont {A.~A.}\ \bibnamefont
  {Balandin}}, \bibinfo {author} {\bibfnamefont {S.}~\bibnamefont {Ghosh}},
  \bibinfo {author} {\bibfnamefont {W.}~\bibnamefont {Bao}}, \bibinfo {author}
  {\bibfnamefont {I.}~\bibnamefont {Calizo}}, \bibinfo {author} {\bibfnamefont
  {D.}~\bibnamefont {Teweldebrhan}}, \bibinfo {author} {\bibfnamefont
  {F.}~\bibnamefont {Miao}}, \ and\ \bibinfo {author} {\bibfnamefont {C.~N.}\
  \bibnamefont {Lau}},\ }\href@noop {} {\bibfield  {journal} {\bibinfo
  {journal} {Nano Lett.}\ }\textbf {\bibinfo {volume} {8}},\ \bibinfo {pages}
  {902} (\bibinfo {year} {2008})}\BibitemShut {NoStop}%
\bibitem [{\citenamefont {Hwang}\ and\ \citenamefont
  {Sarma}(2007)}]{hwang2007PRB}%
  \BibitemOpen
  \bibfield  {author} {\bibinfo {author} {\bibfnamefont {E.}~\bibnamefont
  {Hwang}}\ and\ \bibinfo {author} {\bibfnamefont {S.~D.}\ \bibnamefont
  {Sarma}},\ }\href@noop {} {\bibfield  {journal} {\bibinfo  {journal} {Phys.
  Rev. B}\ }\textbf {\bibinfo {volume} {75}},\ \bibinfo {pages} {205418}
  (\bibinfo {year} {2007})}\BibitemShut {NoStop}%
\bibitem [{\citenamefont {Otsuji}\ \emph {et~al.}(2014)\citenamefont {Otsuji},
  \citenamefont {Popov},\ and\ \citenamefont {Ryzhii}}]{otsuji2014JPhysD}%
  \BibitemOpen
  \bibfield  {author} {\bibinfo {author} {\bibfnamefont {T.}~\bibnamefont
  {Otsuji}}, \bibinfo {author} {\bibfnamefont {V.}~\bibnamefont {Popov}}, \
  and\ \bibinfo {author} {\bibfnamefont {V.}~\bibnamefont {Ryzhii}},\
  }\href@noop {} {\bibfield  {journal} {\bibinfo  {journal} {J. Phys. D: Appl.
  Phys.}\ }\textbf {\bibinfo {volume} {47}},\ \bibinfo {pages} {094006}
  (\bibinfo {year} {2014})}\BibitemShut {NoStop}%
\bibitem [{\citenamefont {Berman}\ \emph {et~al.}(2013)\citenamefont {Berman},
  \citenamefont {Kezerashvili},\ and\ \citenamefont {Lozovik}}]{berman2013PRB}%
  \BibitemOpen
  \bibfield  {author} {\bibinfo {author} {\bibfnamefont {O.~L.}\ \bibnamefont
  {Berman}}, \bibinfo {author} {\bibfnamefont {R.~Y.}\ \bibnamefont
  {Kezerashvili}}, \ and\ \bibinfo {author} {\bibfnamefont {Y.~E.}\
  \bibnamefont {Lozovik}},\ }\href@noop {} {\bibfield  {journal} {\bibinfo
  {journal} {Physical Review B}\ }\textbf {\bibinfo {volume} {88}},\ \bibinfo
  {pages} {235424} (\bibinfo {year} {2013})}\BibitemShut {NoStop}%
\bibitem [{\citenamefont {Rupasinghe}\ \emph {et~al.}(2014)\citenamefont
  {Rupasinghe}, \citenamefont {Rukhlenko},\ and\ \citenamefont
  {Premaratne}}]{rupasinghe2014ACSNano}%
  \BibitemOpen
  \bibfield  {author} {\bibinfo {author} {\bibfnamefont {C.}~\bibnamefont
  {Rupasinghe}}, \bibinfo {author} {\bibfnamefont {I.~D.}\ \bibnamefont
  {Rukhlenko}}, \ and\ \bibinfo {author} {\bibfnamefont {M.}~\bibnamefont
  {Premaratne}},\ }\href@noop {} {\bibfield  {journal} {\bibinfo  {journal}
  {ACS Nano}\ }\textbf {\bibinfo {volume} {8}},\ \bibinfo {pages} {2431}
  (\bibinfo {year} {2014})}\BibitemShut {NoStop}%
\bibitem [{\citenamefont {Apalkov}\ and\ \citenamefont
  {Stockman}(2014)}]{apalkov2014LSA}%
  \BibitemOpen
  \bibfield  {author} {\bibinfo {author} {\bibfnamefont {V.}~\bibnamefont
  {Apalkov}}\ and\ \bibinfo {author} {\bibfnamefont {M.~I.}\ \bibnamefont
  {Stockman}},\ }\href@noop {} {\bibfield  {journal} {\bibinfo  {journal}
  {Light: Science \& Applications}\ }\textbf {\bibinfo {volume} {3}},\ \bibinfo
  {pages} {e191} (\bibinfo {year} {2014})}\BibitemShut {NoStop}%
\bibitem [{\citenamefont {Carmichael}(2008)}]{carmichael2008book2}%
  \BibitemOpen
  \bibfield  {author} {\bibinfo {author} {\bibfnamefont {H.~J.}\ \bibnamefont
  {Carmichael}},\ }\href {\doibase 10.1007/978-3-540-71320-3} {\emph {\bibinfo
  {title} {Statistical methods in quantum optics}}},\ Vol.~\bibinfo {volume}
  {2}\ (\bibinfo  {publisher} {Springer},\ \bibinfo {year} {2008})\BibitemShut
  {NoStop}%
\bibitem [{\citenamefont {Khurgin}\ and\ \citenamefont
  {Sun}(2012{\natexlab{a}})}]{khurgin2012OE}%
  \BibitemOpen
  \bibfield  {author} {\bibinfo {author} {\bibfnamefont {J.~B.}\ \bibnamefont
  {Khurgin}}\ and\ \bibinfo {author} {\bibfnamefont {G.}~\bibnamefont {Sun}},\
  }\href@noop {} {\bibfield  {journal} {\bibinfo  {journal} {Optics Express}\
  }\textbf {\bibinfo {volume} {20}},\ \bibinfo {pages} {15309} (\bibinfo {year}
  {2012}{\natexlab{a}})}\BibitemShut {NoStop}%
\bibitem [{\citenamefont {Khurgin}\ and\ \citenamefont
  {Sun}(2012{\natexlab{b}})}]{khurgin2012APL}%
  \BibitemOpen
  \bibfield  {author} {\bibinfo {author} {\bibfnamefont {J.~B.}\ \bibnamefont
  {Khurgin}}\ and\ \bibinfo {author} {\bibfnamefont {G.}~\bibnamefont {Sun}},\
  }\href@noop {} {\bibfield  {journal} {\bibinfo  {journal} {Appl. Phys.
  Lett.}\ }\textbf {\bibinfo {volume} {100}},\ \bibinfo {pages} {011105}
  (\bibinfo {year} {2012}{\natexlab{b}})}\BibitemShut {NoStop}%
\bibitem [{\citenamefont {Andrianov}\ \emph {et~al.}(2013)\citenamefont
  {Andrianov}, \citenamefont {Baranov}, \citenamefont {Pukhov}, \citenamefont
  {Dorofeenko}, \citenamefont {Vinogradov},\ and\ \citenamefont
  {Lisyansky}}]{andrianov2013loss}%
  \BibitemOpen
  \bibfield  {author} {\bibinfo {author} {\bibfnamefont {E.}~\bibnamefont
  {Andrianov}}, \bibinfo {author} {\bibfnamefont {D.}~\bibnamefont {Baranov}},
  \bibinfo {author} {\bibfnamefont {A.}~\bibnamefont {Pukhov}}, \bibinfo
  {author} {\bibfnamefont {A.}~\bibnamefont {Dorofeenko}}, \bibinfo {author}
  {\bibfnamefont {A.}~\bibnamefont {Vinogradov}}, \ and\ \bibinfo {author}
  {\bibfnamefont {A.}~\bibnamefont {Lisyansky}},\ }\href@noop {} {\bibfield
  {journal} {\bibinfo  {journal} {Optics express}\ }\textbf {\bibinfo {volume}
  {21}},\ \bibinfo {pages} {13467} (\bibinfo {year} {2013})}\BibitemShut
  {NoStop}%
\bibitem [{\citenamefont {Falkovsky}(2008)}]{falkovsky2008PhysUsp}%
  \BibitemOpen
  \bibfield  {author} {\bibinfo {author} {\bibfnamefont {L.~A.}\ \bibnamefont
  {Falkovsky}},\ }\href@noop {} {\bibfield  {journal} {\bibinfo  {journal}
  {Physics-Uspekhi}\ }\textbf {\bibinfo {volume} {51}},\ \bibinfo {pages} {887}
  (\bibinfo {year} {2008})}\BibitemShut {NoStop}%
\bibitem [{\citenamefont {Kotov}\ \emph {et~al.}(2013)\citenamefont {Kotov},
  \citenamefont {Kol'chenko},\ and\ \citenamefont {Lozovik}}]{kotov2013OptExp}%
  \BibitemOpen
  \bibfield  {author} {\bibinfo {author} {\bibfnamefont {O.}~\bibnamefont
  {Kotov}}, \bibinfo {author} {\bibfnamefont {M.}~\bibnamefont {Kol'chenko}}, \
  and\ \bibinfo {author} {\bibfnamefont {Y.~E.}\ \bibnamefont {Lozovik}},\
  }\href@noop {} {\bibfield  {journal} {\bibinfo  {journal} {Optics express}\
  }\textbf {\bibinfo {volume} {21}},\ \bibinfo {pages} {13533} (\bibinfo {year}
  {2013})}\BibitemShut {NoStop}%
\bibitem [{\citenamefont {Landau}\ \emph {et~al.}(1984)\citenamefont {Landau},
  \citenamefont {Bell}, \citenamefont {Kearsley}, \citenamefont {Pitaevskii},
  \citenamefont {Lifshitz},\ and\ \citenamefont
  {Sykes}}]{landau1984electrodynamics}%
  \BibitemOpen
  \bibfield  {author} {\bibinfo {author} {\bibfnamefont {L.~D.}\ \bibnamefont
  {Landau}}, \bibinfo {author} {\bibfnamefont {J.}~\bibnamefont {Bell}},
  \bibinfo {author} {\bibfnamefont {M.}~\bibnamefont {Kearsley}}, \bibinfo
  {author} {\bibfnamefont {L.}~\bibnamefont {Pitaevskii}}, \bibinfo {author}
  {\bibfnamefont {E.}~\bibnamefont {Lifshitz}}, \ and\ \bibinfo {author}
  {\bibfnamefont {J.}~\bibnamefont {Sykes}},\ }\href@noop {} {\emph {\bibinfo
  {title} {Electrodynamics of continuous media}}},\ Vol.~\bibinfo {volume} {8}\
  (\bibinfo  {publisher} {elsevier},\ \bibinfo {year} {1984})\BibitemShut
  {NoStop}%
\bibitem [{\citenamefont {Novotny}\ and\ \citenamefont
  {Hecht}(2012)}]{novotny2012principles}%
  \BibitemOpen
  \bibfield  {author} {\bibinfo {author} {\bibfnamefont {L.}~\bibnamefont
  {Novotny}}\ and\ \bibinfo {author} {\bibfnamefont {B.}~\bibnamefont
  {Hecht}},\ }\href@noop {} {\emph {\bibinfo {title} {Principles of
  nano-optics}}}\ (\bibinfo  {publisher} {Cambridge university press},\
  \bibinfo {year} {2012})\BibitemShut {NoStop}%
\bibitem [{\citenamefont {Haken}(1983)}]{haken1983laser}%
  \BibitemOpen
  \bibfield  {author} {\bibinfo {author} {\bibfnamefont {H.}~\bibnamefont
  {Haken}},\ }\href@noop {} {\emph {\bibinfo {title} {Laser theory}}}\
  (\bibinfo  {publisher} {Springer-Verlag, New York, NY, USA},\ \bibinfo {year}
  {1983})\BibitemShut {NoStop}%
\bibitem [{\citenamefont {Siegman}(1971)}]{siegman1986lasers}%
  \BibitemOpen
  \bibfield  {author} {\bibinfo {author} {\bibfnamefont {A.~E.}\ \bibnamefont
  {Siegman}},\ }\href@noop {} {\  (\bibinfo {year} {1971})}\BibitemShut
  {NoStop}%
\bibitem [{\citenamefont {Khanin}(2005)}]{khanin2005fundamentals}%
  \BibitemOpen
  \bibfield  {author} {\bibinfo {author} {\bibfnamefont {Y.~I.}\ \bibnamefont
  {Khanin}},\ }\href@noop {} {\emph {\bibinfo {title} {Fundamentals of laser
  dynamics}}}\ (\bibinfo  {publisher} {Cambridge Int Science Publishing},\
  \bibinfo {year} {2005})\BibitemShut {NoStop}%
\bibitem [{\citenamefont {Chow}\ \emph {et~al.}(2014)\citenamefont {Chow},
  \citenamefont {Jahnke},\ and\ \citenamefont {Gies}}]{chow2014LSA}%
  \BibitemOpen
  \bibfield  {author} {\bibinfo {author} {\bibfnamefont {W.~W.}\ \bibnamefont
  {Chow}}, \bibinfo {author} {\bibfnamefont {F.}~\bibnamefont {Jahnke}}, \ and\
  \bibinfo {author} {\bibfnamefont {C.}~\bibnamefont {Gies}},\ }\href@noop {}
  {\bibfield  {journal} {\bibinfo  {journal} {Light Sci. Appl.}\ }\textbf
  {\bibinfo {volume} {3}},\ \bibinfo {pages} {e201} (\bibinfo {year}
  {2014})}\BibitemShut {NoStop}%
\bibitem [{\citenamefont {Protsenko}(2012)}]{protsenko2012theory}%
  \BibitemOpen
  \bibfield  {author} {\bibinfo {author} {\bibfnamefont {I.~E.}\ \bibnamefont
  {Protsenko}},\ }\href@noop {} {\bibfield  {journal} {\bibinfo  {journal}
  {Physics-Uspekhi}\ }\textbf {\bibinfo {volume} {55}},\ \bibinfo {pages}
  {1040} (\bibinfo {year} {2012})}\BibitemShut {NoStop}%
\bibitem [{\citenamefont {Protsenko}\ \emph {et~al.}(1999)\citenamefont
  {Protsenko}, \citenamefont {Domokos}, \citenamefont {Lef{\`e}vre-Seguin},
  \citenamefont {Hare}, \citenamefont {Raimond},\ and\ \citenamefont
  {Davidovich}}]{protsenko1999PRA}%
  \BibitemOpen
  \bibfield  {author} {\bibinfo {author} {\bibfnamefont {I.}~\bibnamefont
  {Protsenko}}, \bibinfo {author} {\bibfnamefont {P.}~\bibnamefont {Domokos}},
  \bibinfo {author} {\bibfnamefont {V.}~\bibnamefont {Lef{\`e}vre-Seguin}},
  \bibinfo {author} {\bibfnamefont {J.}~\bibnamefont {Hare}}, \bibinfo {author}
  {\bibfnamefont {J.}~\bibnamefont {Raimond}}, \ and\ \bibinfo {author}
  {\bibfnamefont {L.}~\bibnamefont {Davidovich}},\ }\href@noop {} {\bibfield
  {journal} {\bibinfo  {journal} {Phys. Rev. A}\ }\textbf {\bibinfo {volume}
  {59}},\ \bibinfo {pages} {1667} (\bibinfo {year} {1999})}\BibitemShut
  {NoStop}%
\bibitem [{\citenamefont {Khurgin}\ and\ \citenamefont
  {Sun}(2014)}]{khurgin2014NatPhot}%
  \BibitemOpen
  \bibfield  {author} {\bibinfo {author} {\bibfnamefont {J.~B.}\ \bibnamefont
  {Khurgin}}\ and\ \bibinfo {author} {\bibfnamefont {G.}~\bibnamefont {Sun}},\
  }\href@noop {} {\bibfield  {journal} {\bibinfo  {journal} {Nat. Photonics}\
  }\textbf {\bibinfo {volume} {8}},\ \bibinfo {pages} {468} (\bibinfo {year}
  {2014})}\BibitemShut {NoStop}%
\bibitem [{\citenamefont {Pietryga}\ \emph {et~al.}(2004)\citenamefont
  {Pietryga}, \citenamefont {Schaller}, \citenamefont {Werder}, \citenamefont
  {Stewart}, \citenamefont {Klimov},\ and\ \citenamefont
  {Hollingsworth}}]{pietryga2004pushing}%
  \BibitemOpen
  \bibfield  {author} {\bibinfo {author} {\bibfnamefont {J.~M.}\ \bibnamefont
  {Pietryga}}, \bibinfo {author} {\bibfnamefont {R.~D.}\ \bibnamefont
  {Schaller}}, \bibinfo {author} {\bibfnamefont {D.}~\bibnamefont {Werder}},
  \bibinfo {author} {\bibfnamefont {M.~H.}\ \bibnamefont {Stewart}}, \bibinfo
  {author} {\bibfnamefont {V.~I.}\ \bibnamefont {Klimov}}, \ and\ \bibinfo
  {author} {\bibfnamefont {J.~A.}\ \bibnamefont {Hollingsworth}},\ }\href@noop
  {} {\bibfield  {journal} {\bibinfo  {journal} {Journal of the American
  Chemical Society}\ }\textbf {\bibinfo {volume} {126}},\ \bibinfo {pages}
  {11752} (\bibinfo {year} {2004})}\BibitemShut {NoStop}%
\bibitem [{\citenamefont {Wehrenberg}\ \emph {et~al.}(2002)\citenamefont
  {Wehrenberg}, \citenamefont {Wang},\ and\ \citenamefont
  {Guyot-Sionnest}}]{wehrenberg2002JCP}%
  \BibitemOpen
  \bibfield  {author} {\bibinfo {author} {\bibfnamefont {B.~L.}\ \bibnamefont
  {Wehrenberg}}, \bibinfo {author} {\bibfnamefont {C.}~\bibnamefont {Wang}}, \
  and\ \bibinfo {author} {\bibfnamefont {P.}~\bibnamefont {Guyot-Sionnest}},\
  }\href@noop {} {\bibfield  {journal} {\bibinfo  {journal} {J. Phys. Chem. B}\
  }\textbf {\bibinfo {volume} {106}},\ \bibinfo {pages} {10634} (\bibinfo
  {year} {2002})}\BibitemShut {NoStop}%
\bibitem [{\citenamefont {Keuleyan}\ \emph {et~al.}(2011)\citenamefont
  {Keuleyan}, \citenamefont {Lhuillier},\ and\ \citenamefont
  {Guyot-Sionnest}}]{keuleyan2011synthesis}%
  \BibitemOpen
  \bibfield  {author} {\bibinfo {author} {\bibfnamefont {S.}~\bibnamefont
  {Keuleyan}}, \bibinfo {author} {\bibfnamefont {E.}~\bibnamefont {Lhuillier}},
  \ and\ \bibinfo {author} {\bibfnamefont {P.}~\bibnamefont {Guyot-Sionnest}},\
  }\href@noop {} {\bibfield  {journal} {\bibinfo  {journal} {Journal of the
  American Chemical Society}\ }\textbf {\bibinfo {volume} {133}},\ \bibinfo
  {pages} {16422} (\bibinfo {year} {2011})}\BibitemShut {NoStop}%
\bibitem [{\citenamefont {Lhuillier}\ \emph {et~al.}(2013)\citenamefont
  {Lhuillier}, \citenamefont {Keuleyan}, \citenamefont {Liu},\ and\
  \citenamefont {Guyot-Sionnest}}]{lhuiller2013ChemMat}%
  \BibitemOpen
  \bibfield  {author} {\bibinfo {author} {\bibfnamefont {E.}~\bibnamefont
  {Lhuillier}}, \bibinfo {author} {\bibfnamefont {S.}~\bibnamefont {Keuleyan}},
  \bibinfo {author} {\bibfnamefont {H.}~\bibnamefont {Liu}}, \ and\ \bibinfo
  {author} {\bibfnamefont {P.}~\bibnamefont {Guyot-Sionnest}},\ }\href@noop {}
  {\bibfield  {journal} {\bibinfo  {journal} {Chem. Mater.}\ }\textbf {\bibinfo
  {volume} {25}},\ \bibinfo {pages} {1272} (\bibinfo {year}
  {2013})}\BibitemShut {NoStop}%
\bibitem [{\citenamefont {Mirov}\ \emph {et~al.}(2010)\citenamefont {Mirov},
  \citenamefont {Fedorov}, \citenamefont {Moskalev}, \citenamefont
  {Martyshkin},\ and\ \citenamefont {Kim}}]{mirov2010LPR}%
  \BibitemOpen
  \bibfield  {author} {\bibinfo {author} {\bibfnamefont {S.~B.}\ \bibnamefont
  {Mirov}}, \bibinfo {author} {\bibfnamefont {V.~V.}\ \bibnamefont {Fedorov}},
  \bibinfo {author} {\bibfnamefont {I.~S.}\ \bibnamefont {Moskalev}}, \bibinfo
  {author} {\bibfnamefont {D.~V.}\ \bibnamefont {Martyshkin}}, \ and\ \bibinfo
  {author} {\bibfnamefont {S.}~\bibnamefont {Kim}},\ }\href@noop {} {\bibfield
  {journal} {\bibinfo  {journal} {Laser and Photon. Rev.}\ }\textbf {\bibinfo
  {volume} {4}},\ \bibinfo {pages} {21} (\bibinfo {year} {2010})}\BibitemShut
  {NoStop}%
\bibitem [{\citenamefont {Wang}\ \emph {et~al.}(2007)\citenamefont {Wang},
  \citenamefont {Cha}, \citenamefont {Liu},\ and\ \citenamefont
  {Chen}}]{wang2007IEEE}%
  \BibitemOpen
  \bibfield  {author} {\bibinfo {author} {\bibfnamefont {K.~L.}\ \bibnamefont
  {Wang}}, \bibinfo {author} {\bibfnamefont {D.}~\bibnamefont {Cha}}, \bibinfo
  {author} {\bibfnamefont {J.}~\bibnamefont {Liu}}, \ and\ \bibinfo {author}
  {\bibfnamefont {C.}~\bibnamefont {Chen}},\ }\href@noop {} {\bibfield
  {journal} {\bibinfo  {journal} {Proceedings of the IEEE}\ }\textbf {\bibinfo
  {volume} {95}},\ \bibinfo {pages} {1866} (\bibinfo {year}
  {2007})}\BibitemShut {NoStop}%
\bibitem [{\citenamefont {Botez}(2007)}]{botez2007INJ}%
  \BibitemOpen
  \bibfield  {author} {\bibinfo {author} {\bibfnamefont {D.}~\bibnamefont
  {Botez}},\ }\href@noop {} {\bibfield  {journal} {\bibinfo  {journal}
  {International Journal of Nanoscience}\ }\textbf {\bibinfo {volume} {6}},\
  \bibinfo {pages} {203} (\bibinfo {year} {2007})}\BibitemShut {NoStop}%
\bibitem [{\citenamefont {Mirov}\ \emph {et~al.}(2007)\citenamefont {Mirov},
  \citenamefont {Fedorov}, \citenamefont {Moskalev},\ and\ \citenamefont
  {Martyshkin}}]{mirov2007JSTQE}%
  \BibitemOpen
  \bibfield  {author} {\bibinfo {author} {\bibfnamefont {S.~B.}\ \bibnamefont
  {Mirov}}, \bibinfo {author} {\bibfnamefont {V.~V.}\ \bibnamefont {Fedorov}},
  \bibinfo {author} {\bibfnamefont {I.~S.}\ \bibnamefont {Moskalev}}, \ and\
  \bibinfo {author} {\bibfnamefont {D.~V.}\ \bibnamefont {Martyshkin}},\
  }\href@noop {} {\bibfield  {journal} {\bibinfo  {journal} {IEEE Journal of
  Selected Topics in Quantum Electronics}\ }\textbf {\bibinfo {volume} {13}},\
  \bibinfo {pages} {810} (\bibinfo {year} {2007})}\BibitemShut {NoStop}%
\end{thebibliography}%

\end{document}